\documentclass{article}

\usepackage{amssymb}
\usepackage{amsmath}
\usepackage{caption}
\usepackage{subfigure}
\usepackage{multicol}
\usepackage{here}
\usepackage{graphicx}
\usepackage{color}
\usepackage{algorithmicx}
\usepackage{algpseudocode}
\usepackage{algorithm}
\usepackage{listings}
\usepackage{epstopdf}
\usepackage{marvosym}
\usepackage{authblk}

\DeclareGraphicsExtensions{.eps}
\def\calV{\mathcal{V}}
\def\fatA{\mathbf{A}}

\def\fatD{\mathbf{D}}

\def\fatE{\mathbf{E}}

\def\fatp{\mathbf{p}}

\def\fatn{\mathbf{n}}

\def\fatu{\mathbf{u}}
\def\fatx{\mathbf{x}}

\def\fatK{\mathbf{K}}

\def\fatx{\mathbf{x}}
\def\faty{\mathbf{y}}

\def\fatmu{\boldsymbol{\mu}}

\numberwithin{equation}{section}
\numberwithin{table}{section}
\numberwithin{figure}{section}

\begin{document}

\title{Multiscale modeling of diffusion in a crowded environment}
\author{Lina Meinecke\thanks{lina.meinecke@it.uu.se}}
\affil{Department of Information Technology, Uppsala University}
\maketitle

\begin{abstract}
  We present a multiscale approach to model diffusion in a crowded environment and its effect on the reaction rates.
  Diffusion in biological systems is often modeled by a discrete space jump process in order to capture the inherent noise of biological systems, which becomes important in the low copy number regime.
  To model diffusion in the crowded cell environment efficiently, we compute the jump rates in this mesoscopic model from local first exit times, which account for the microscopic positions of the crowding molecules, while the diffusing molecules jump on a coarser Cartesian grid.
  We then extract a macroscopic description from the resulting jump rates, where the excluded volume effect is modeled by a diffusion equation with space dependent diffusion coefficient.
  The crowding molecules can be of arbitrary shape and size and numerical experiments demonstrate that those factors together with the size of the diffusing molecule play a crucial role on the magnitude of the decrease in diffusive motion.
  When correcting the reaction rates for the altered diffusion we can show that molecular crowding either enhances or inhibits chemical reactions depending on local fluctuations of the obstacle density.
\end{abstract}


\section{Introduction}\label{sec:intro}
Living cells are spatially organized, e.g. eukaryotic cells have a confined nucleus containing the DNA and reaction complexes are often bound to the cell membrane.
To simulate reaction networks accurately it therefore is important to incorporate the molecules' movement into the models and account for the time it takes for a signal to transmit e.g. from the nucleus to the membrane.

Molecules move by diffusion through biological media such as the cytoplasm, which is a non-solute medium, where an estimated $40\%$ \cite{Luby-Phelps2000,Schnell2004} of the available space is occupied by macromolecules, such as proteins, ribosomes, RNA and the cytoskeleton.
The environment is called crowded, meaning that the space is densely packed by molecules but individual species are only present at very low concentrations.
Macromolecular crowding is especially important on the cell membrane \cite{Grasberger1986}, where attaching actin filaments \cite{Medalia2002} create barriers, that hinder the displacement of membrane bound molecules \cite{Jin2007,Krapf2015}.
In mitochondria more than $60\%$ of the matrix can be occupied by enzymes and proteins \cite{Verkman2002} and also the extracellular space between e.g. brain cells \cite{Hrabe2004} is considered crowded.

The steric repulsions between molecules in a crowded environment force diffusing molecules to move around obstacles, or "crowders", this slows down diffusion.
New techniques such as fluorescence-fluctuation analysis \cite{DiRienzo2014} have shown that diffusion is not simply slowed down but that crowding can lead to anomalous diffusion, where the mean square displacement (MSD) of the molecule is no longer linear, but sublinear in time.
As the crowder density increases, space is divided into subdomains and becomes inhomogeneous. 
For this fractal space, the dimension decreases to a non integer and the MSD no longer follows the linear law applicable in integer dimensions \cite{Ben-Avraham2000,Havlin2002}.

The change of the diffusion rate in a crowded environment is a hydrodynamic effect. 
The excluded volume effect on the reaction rates is a thermodynamic consequence \cite{Hall2003} and can be both impeding and promoting.
While diffusion limited reaction rates are decreased due to the slower diffusion, transition state limited reactions and dimerizations are accelerated \cite{Ellis2001} since intermediate products reside longer in the vicinity of reaction complexes and dimers occupy less volume than two monomers.
Hindered diffusion also leads to localized reactions and a heterogeneous distribution of products, which increases intrinsic noise \cite{Hansen2015}.

Scaled particle theory (SPT) has been used to describe the thermodynamic effect on the reaction rates in a crowded envrionment \cite{Grima2010, Hall2003, Ridgway2008}.
Another approach is to perform Brownian dynamics (BD) simulations and fit the reaction rates to the microscopic results, \cite{Lee2008,Smith2014}.
In \cite{Berry2002}, Michaelis-Menten reaction dynamics are best fitted by fractal kinetics and the results are verified by microscopic cellular automata (CA) simulations.
The fractal kinetics are modified in \cite{Schnell2004} to a Zipf-Mandelbrot distribution of the reaction rates.

To better understand the effects of excluded volume on both diffusion and reactions, accurate reaction-diffusion simulations in the crowded cell environment are needed. 
The microscopic approaches mentioned above are computationally very expensive due to the high number of collisions in such a medium.
In this paper we present a novel multiscale approach to simulate diffusion of a spherical particle surrounded by inert and inactive crowders of any size and shape.
We resolve the microscopic positions and shapes of the crowders initially to precompute jump rates for the moving molecules.
The molecule follows a random walk on a coarse Cartesian grid that no longer resolves the multiple crowders for computationally more efficient simulations. 
With our approach we can connect a given distribution of obstacles to a space dependent diffusion map which can be used to recompute space dependent reaction rates representing reactions in the crowded environment.
The method can easily be extended to moving crowders and an advantage over other techniques such as SPT is the versatility in the shape of the crowders.
The upscaling to a coarse grid makes the stochastic simulations computationally much more efficient than BD and CA simulations.

In the next section we present existing models of spatial simulations in systems biology and how they incorporate crowding effects.
We then present how the microscopic motion of a molecule can be used to calculate its first exit time (FET) from domains, which provides the jump rates in a coarse grained discrete jump process on the mesoscopic level.
We continue by extending the FET approach to include macromolecular crowders.
In Section \ref{sec:macro}, we use the jump coefficients and compute a space dependent diffusion map for the macroscopic level and show in Section \ref{sec:reactions} how that affects the reaction rates in the crowded environment.
We conclude with numerical experiments in the final section.

Vectors and matrices are written in boldface. A vector $\fatu$ has the
components $u_i$ and the elements of a matrix $\fatA$ are
$A_{ij}$. The derivative of a variable $u$ with respect to time $t$ is
written $u_t$. 


\section{Spatial modeling in systems biology}\label{sec:Spatial}
In this section we first present existing models of diffusion simulations in systems biology and then describe how they can be adapted to include macromolecular crowding.

\subsection{Models of diffusion in dilute media}
Molecules undergoing diffusion and reactions inside living cells are often modeled by the reaction-diffusion equations.
These are continuous, deterministic partial differential equations (PDEs) describing the time evolution of the concentrations of molecules.
For a diffusing molecule the concentration $u(\fatx,t)$ is described by the diffusion equation
\begin{equation}\label{eq:DiffFree}
u_t(\fatx,t) = \gamma_0\Delta u(\fatx,t),\quad\fatx\in\Omega
\end{equation}
with diffusion coefficient $\gamma_0$ and suitable boundary conditions on $\partial\Omega$.
To include reactions corresponding terms are added.
This \textbf{macroscopic} description is accurate in the limit of large molecule numbers, when stochastic fluctuations are small and the mean value is the quantity of interest.
Important molecules such as DNA or transcription factors are, however, only present at very low copy numbers inside living cells.
It has been observed in experiments \cite{ELSS, MAA1, Met, MunskyNeuertOuden, RajOuden, Swain01102002} and shown theoretically \cite{GaMcWaMa, McQuarrie} that stochastic fluctuations play an important role and a discrete stochastic description is more accurate than the deterministic equations.

We distinguish two levels of accuracy of stochastic models.
In the \textbf{mesoscopic} model the domain $\Omega$ is partitioned into $N$ non-overlapping voxels $\calV_i$ with nodes $\fatx_i$ at the center.
The state vector $\faty(t)$ contains the number of molecules $y_i(t)$ in each voxel $\calV_i$ at time $t$.
The voxels are small enough that the molecules can be considered well mixed inside so that reactions can occur between molecules residing in the same voxel.
An individual molecule can jump from a voxel $\calV_i$ to a neighboring voxel $\calV_j$ to model diffusion. 
The diffusion master equation (DME) describes the time evolution of the probability to be in state $\faty$ of a system with only diffusion 
\begin{equation}
\frac{\partial p(\faty,t)}{\partial t} = \sum_{i=1}^N\sum_{j=1}^N\lambda_{ij}(\faty-\fatmu_{ij})p(\faty-\fatmu_{ij},t)-\lambda_{ii}(\faty)p(\faty,t),
\end{equation}
where $\lambda_{ij}$ is the jump propensity from $\calV_i$ to $\calV_j$ and $\lambda_{ii}=\sum_{j=1}^N\lambda_{ij}$ is the total propensity to leave voxel $\calV_i$. 
The transition vector $\fatmu_{ij}$ is zero except for $\fatmu_{ij,i}=-1$ and $\fatmu_{ij,j}=1$.
Let $\theta_{ij}$ be the splitting probability, that a jump from $\calV_i$ goes to $\calV_j$, then
\begin{equation}
\lambda_{ij} = \theta_{ij}\lambda_{ii}.
\end{equation}
By including reaction terms in a similar manner, the DME can be extended to the reaction-diffusion master equation (RDME).
In the presence of bimolecular reactions there exists no analytical solution and a numerical solution is impossible due to the high dimension of $\faty$.
Instead, one samples trajectories of the system with the stochastic simulation algorithm (SSA), first presented by Gillespie \cite{gillespie} for only reactions and improved in \cite{CaoGilPet1, GibsonBruck}.
The algorithm was extended to space dependent problems with a Cartesian partioning of the domain in \cite{ElEh04} implemented in \cite{HFE}. 
The propensities $\lambda_{ii}$ 
are here used to generate random numbers for the time until the next jump.
To represent the complicated geometries present in cells the algorithm was extended to curved boundaries in \cite{IsaacsonPeskin} and adapted for unstructured meshes in \cite{EnFeHeLo} with software in \cite{URDME, STEPS}.   

In the more accurate \textbf{microscopic} model the molecules are tracked along their Brownian trajectories in a continuous space, continuous time Markov process.
One approach is called Green's function reaction dynamics (GFRD), methods and software for this approach are found in \cite{AnAdBrAr10, DBOGSK, MCell08, ZoWo5a} with a review in \cite{Schoneberg2014}.
In \cite{Donev2010a,OBDKGS,Takahashi2010} protective domains are constructed around individual molecules in which they cannot interact with other molecules. 
The exit times and exit positions from these domains are sampled to propagate the system until molecules are close enough to interact without sampling all the intermediate jumps.

\subsection{Include macromolecular crowding}\label{sec:Crowding}
The macroscopic, mesoscopic and microscopic models presented above are designed to simulate diffusion in a dilute medium. 
The microscopic model incorporates crowding effects automatically since the molecules are modeled as hard spheres with a given volume, but it becomes computationally very expensive in a densely packed space of inactive crowders, because the protective domains around molecules will be small and many short jumps will be simulated before meeting a potential reaction partner.

Cellular automata (CA) have been used in \cite{Berry2002,Cianci2015, Schnell2004, Takahashi2005} to simulate diffusion in a crowded environment. 
This is a lattice, or voxel, based microscopic approach, where each site can hold one molecule and crowders are represented as already occupied lattice points.
The jump length is here the size of a molecule which also leads to expensive simulations with many short jumps and the shape of the molecules corresponds to the shape of the chosen lattice.
The choice of the lattice, moreover, influences the excluded volume effect \cite{Grima2006}. 
A Cartesian grid leads to a stronger crowding effect than a hexagonal mesh in 2D and both overestimate the effect of crowding on the reaction rates compared to a BD approach.

In \cite{Roberts2013} a mesoscopic approach is used where each voxel can hold more than one molecule.
After distributing immobile crowders it is decided which voxels are accessible and which are full.
The crowders can move in \cite{Fanelli2010,Taylor2015} and the jump propensity to an adjacent node is scaled by the number of available spaces in the target voxel. 
Macroscopic nonlinear PDEs are then derived in \cite{Fanelli2010} to model diffusion in the crowded cell envrionment and the results are validated by physical experiments in \cite{Fanelli2013}.
This approach is extended in \cite{Penington2011} to derive nonlinear diffusion equations modeling more complicated interactions than steric repulsion between the molecules.
Similarily, the averaged occupied volume in the whole domain is used in \cite{Landman2011} to rescale the jump propensities.
These approaches at most take the averaged occupied volume in the target voxel into account and neglect the microscopic positions, the shape of the molecules and the surrounding medium. 
Hence only an averaged behavior is observed and the MSD is linear like for normal diffusion but with a reduced diffusion constant:
\begin{equation}
\langle \fatx^2(t)\rangle = (1-\phi)\gamma_0 t,
\end{equation}
where $\phi$ is the  occupied volume fraction.

In this paper we present a novel multiscale approach to simulate molecular crowding.
We will use the microscopic information of the crowders' positions to recalculate the jump propensities on an overlying mesoscopic mesh.
Instead of simulating diffusion in the detailed environment, as in BD, we will have to solve many PDEs on small subdomains resolving the microscopic positions of the crowders.
This is similar to homogenization techniques used e.g. for flow simulations in porous media, \cite{Brown2014,Malqvist2014}.
The crowding molecules can have any shape and this approach is especially useful when the crowded environment is stationary or evolves on a much slower time scale than the diffusing molecule, so that the jump coefficients can be precomputed and used for a long simulation time.
This is reasonable since the macromolecules responsible for the majority of occupied volume are ribosomes, microtubules and actin filaments \cite{Ellis2001}, which are large and hence diffuse on a slower time scale than for example transcription factors.
Moreover, it was shown in \cite{DiRienzo2014}  that anomalous behavior is most likely to happen in a stationary environment, since otherwise averaging effects simply reduce the coefficient of normal diffusion.
But, it is important to mention that by computing statistics our method can be made efficient also for moving crowders.



\section{Microscopic to mesoscopic: first exit times}\label{sec:FET}

In this section we will use the methods developed for microscopic simulations of Brownian motion with protective domains \cite{OBDKGS}, to derive the jump propensities $\lambda_{ii}$ and splitting probabilities $\theta_{ij}$ for the mesoscopic model.
For simplicity the illustrations are given in dimension $d=2$ but the method can be extended to 3D without modification.

\subsection{First exit times}
Let $c(\fatx,t)$  be the probability distribution that a molecule in Brownian motion is at $\fatx$ at time $t$ and has not yet exited a domain $\omega$. If $\fatx_0$ is the starting position of the molecule diffusing with $\gamma_0$, then $c(\fatx,t)$ fulfills
\begin{align}\label{eq:Diff}
c_t(\fatx,t)  &= \gamma_0\Delta c(\fatx,t), &\fatx & \in\omega,\\
c(\fatx,t) &= 0, &\fatx & \in\partial\omega,\nonumber\\
c(\fatx,0) &= \delta_{\fatx_0}.\nonumber
\end{align}
The homogeneous Dirichlet boundary condition here models that the particle is removed once it hits the boundary.
The survival probability of the particle inside $\omega$ until time $t$ is then
\begin{eqnarray}
S(t) &=& \int_{\omega}c(\fatx,t)d\omega.
\label{eq:S}
\end{eqnarray}
By Gauss' formula the probability density $p_{\omega}(t)$ that the particle leaves $\omega$ at $t$ is
\begin{equation}
p_{\omega}(t) = -\frac{\partial S(t)}{\partial t} = -\gamma_0 \int_{\partial\omega}\fatn\cdot\nabla c(\fatx,t)ds,
\end{equation}
where $\fatn$ is the outward normal.
The expected time $E$ for the molecule to leave $\omega$ for the first time is given by
\begin{equation}
E = \int_0^\infty tp_{\omega}(t)dt = \int_0^\infty S(t)dt.
\label{eq:E}
\end{equation}
We now use this FET approach to compute the jump propensities $\lambda_{ii}$ in the space discrete mesoscopic model.
We use a Cartesian grid with space discretization $h$ and $N$ nodes $\fatx_i$ in the domain $\Omega$. 
The voxels $\calV_i$ are here defined by the dual mesh, see Fig.~\ref{fig:Cartesian}(b).
Since the molecules are considered well mixed inside the voxels, the domain $\omega$ that diffusing particles have to leave to be well mixed in the next voxel has to include the centers of the neighboring voxels. 
On a Cartesian grid we showed in \cite{Lotstedt2015} that solving \eqref{eq:Diff} on the circle $\omega_i$ with center $\fatx_i$ and radius $h$ (similarly a line of length $2h$ in 1D or a sphere with radius $h$ in 3D) and choosing $\fatx_0 = \fatx_i$ gives the correct exit time from node $i$, see Fig.~\ref{fig:Cartesian}(b). 
Observe that $\omega_i\supsetneq\calV_i$ and $\omega_i\cap\omega_j\neq\emptyset$ for neighboring nodes $i$ and $j$.
Using \eqref{eq:S} and \eqref{eq:E} the expected exit time $E_i$ from $\calV_i$ is
\begin{eqnarray}
E_i &=& \frac{h^2}{2d\gamma_0},\label{eq:EMeso}
\end{eqnarray}
see \cite{Gardiner}.
Since the jump propensity is the inverse of the exit time this agrees with the mesoscopic rate on Cartesian grids
\begin{equation}\label{eq:Lambdaii}
\lambda_{ii}=\frac{2d\gamma_0}{h^2}=E_i^{-1}.
\end{equation}
The probability to leave $\omega_i$ through a given part of the boundary $\partial\omega_{ij}$ at time $t$ is given by the proportion of fluxes
\begin{eqnarray}
\theta_{ij}(t) &=&\frac{\int_{\partial\omega_{ij}}\fatn\cdot\nabla c(\fatx,t)ds}{\int_{\partial\omega_i}\fatn\cdot\nabla c(\fatx,t)ds},
\end{eqnarray}
and we can compute the expected probability to jump to a certain neighboring voxel by
\begin{equation}
\theta_{ij} = \int_0^\infty \theta_{ij}(t)p_{\omega_i}(t) dt = -\gamma_0\int_0^t\int_{\partial\omega_{ij}}\fatn\cdot\nabla c(\fatx,t)dsdt.
\end{equation}
Choosing $\partial\omega_{ij}$ to be the quarter segment of the boundary closest to $\fatx_j$, see the blue line in Fig.~\ref{fig:Cartesian}(b), yields the splitting probability $\theta_{ij}=0.25$ as expected on a Cartesian grid.
The method has been extended to a rectangular grid and possible jumps to the diagonal neighbors in \cite{Meinecke2016}.

By conditioning on the first step, these time dependent equations can be converted to equations describing directly the expected quantities \cite{Redner}.
The expected exit time for a molecule starting to diffuse in $\fatx$ from the domain $\omega_i$ fulfills the Poisson's equation
\begin{eqnarray}\label{eq:Poiss}
\gamma_0\Delta E(\fatx)  &= -1, &\quad \fatx\in\omega_i,\\
E(\fatx) &= 0, &\quad \fatx\in\partial\omega_i.\nonumber
\end{eqnarray}
Equivalently, the expected splitting probability for this molecule to exit through the boundary $\partial\omega_{ij}$ can be computed by the harmonic measure \cite[Ch. 7]{Oksendal}, and fulfills the Laplace equation
\begin{eqnarray}\label{eq:Laplace}
\Delta\theta_{ij}(\fatx)  &= 0, &\quad \fatx\in\omega,\\
\theta_{ij}(\fatx) &= 1, &\quad \fatx\in\partial\omega_{ij},\nonumber\\
\theta_{ij}(\fatx) &= 0, &\quad \fatx\in\partial\omega_i\setminus\partial\omega_{ij}.\nonumber
\end{eqnarray}
We solve equations \eqref{eq:Poiss} and \eqref{eq:Laplace} and evaluate them at $\fatx_i$ instead of solving the time-dependent equations and the integrals above.
It is sufficient to solve \eqref{eq:Laplace} three times for each node since $\sum_{j=1}^4\theta_{ij}=1$.

In the following, we will not simply use the circle $\omega_i$ to compute the mesoscopic rates, but we will prohibit the molecule from diffusing where the crowders are located.

\subsection{Include crowding molecules}
\begin{figure}[H]
\subfigure[]{\includegraphics[width=.32\textwidth]{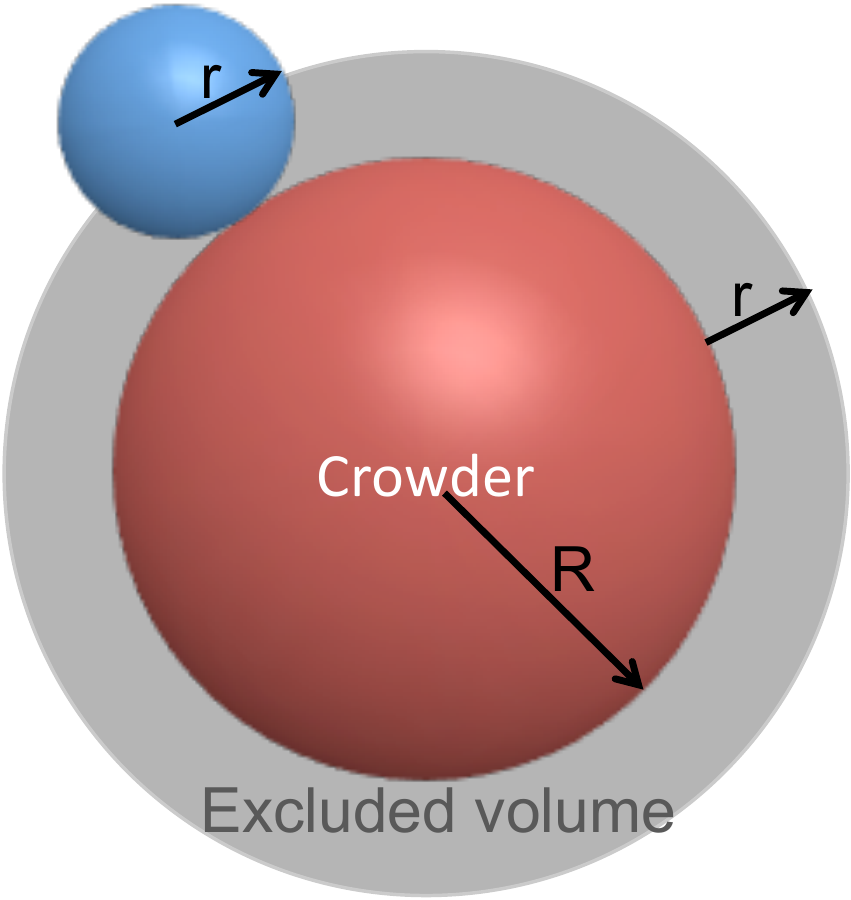} }
\subfigure[]{\includegraphics[width=0.32\textwidth]{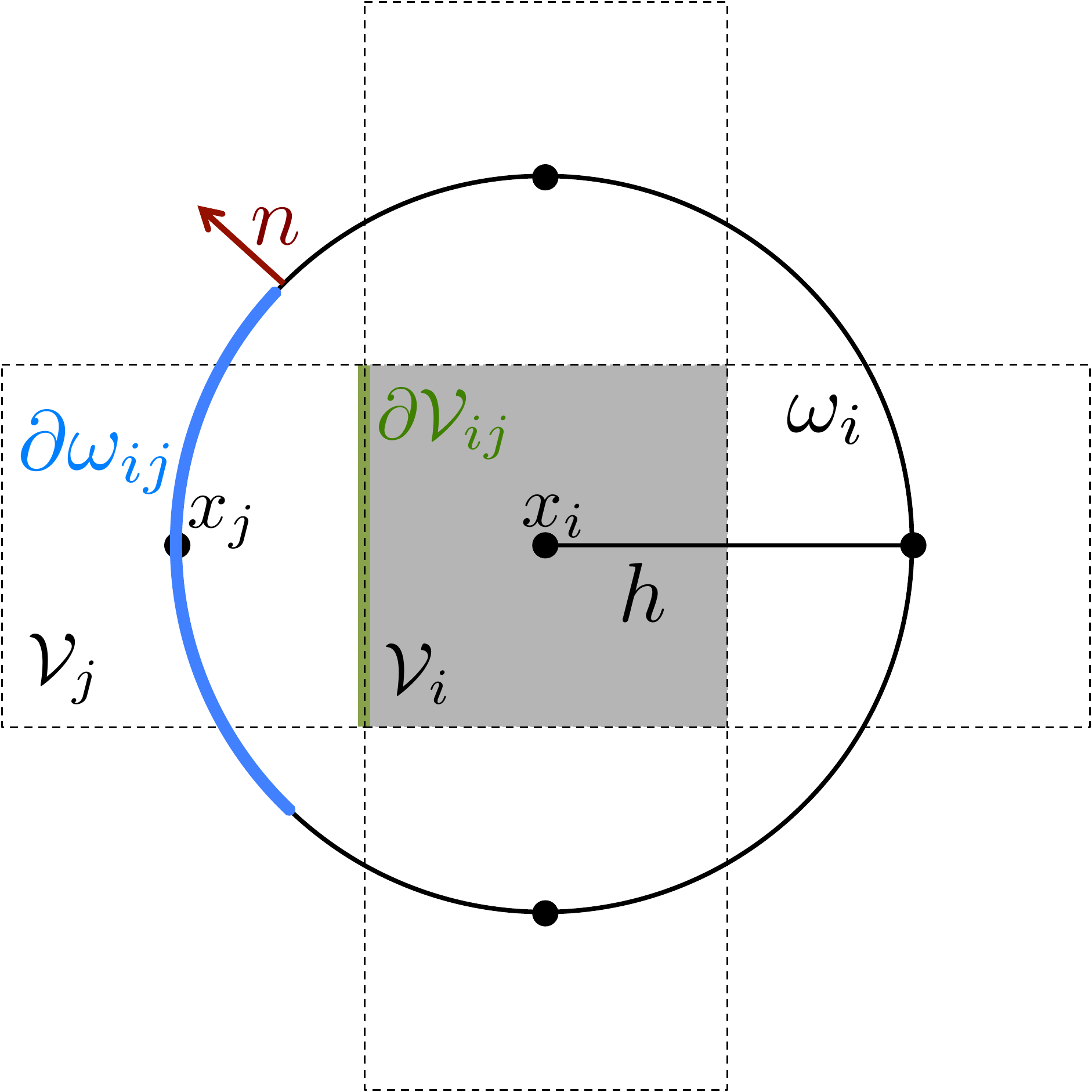}}
\subfigure[]{\includegraphics[width=0.32\textwidth]{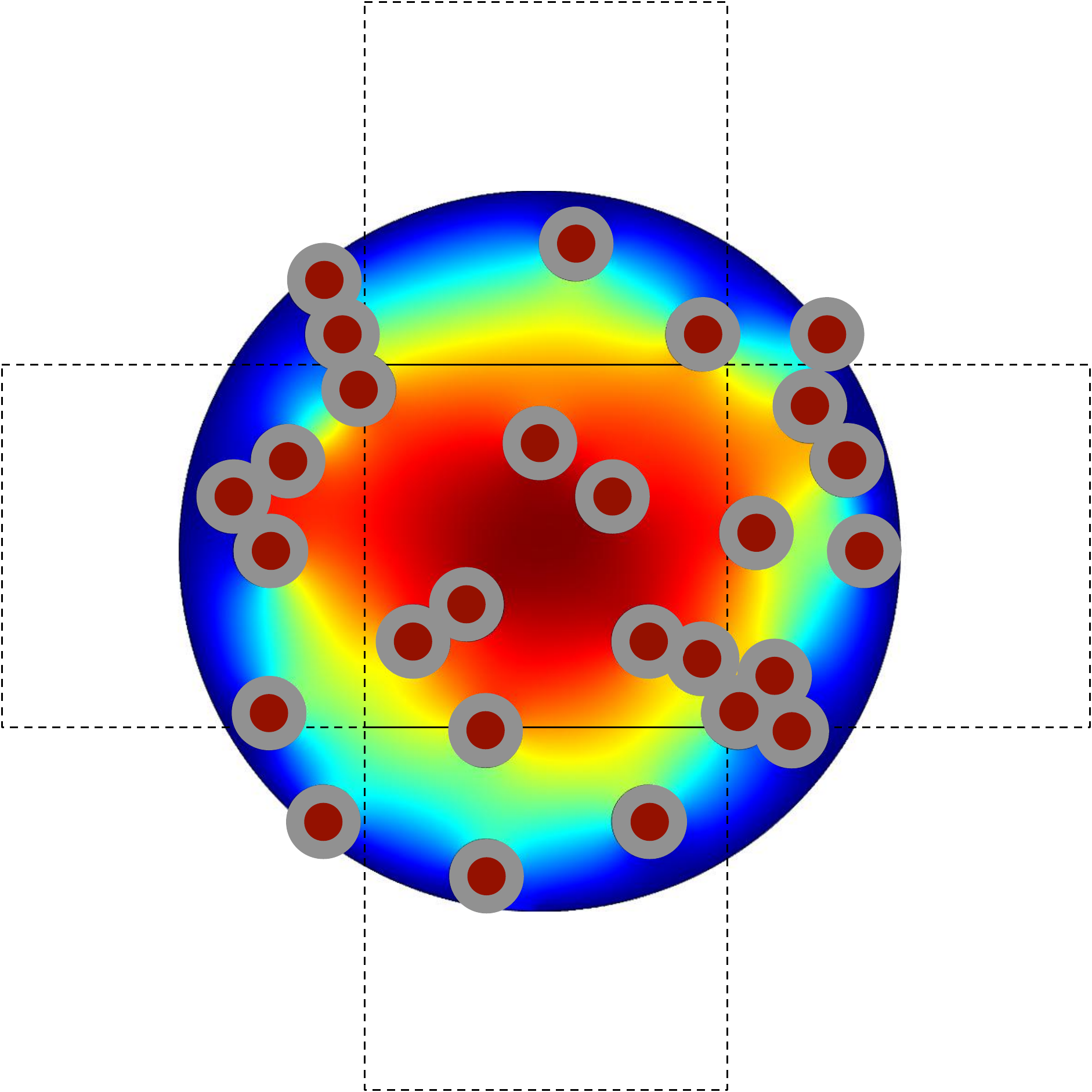}}
\caption{(a)Excluded volume (grey and red) for the center of mass of the diffusing molecule (blue). (b) Cartesian mesh and protective domain $\omega_i$ without crowding. (c) Solution to \eqref{eq:Poiss} under molecular crowding.}\label{fig:Cartesian}
\end{figure}

The crowding molecules are represented as obstacles or holes in the domain $\omega_i$ with reflecting boundary conditions.
Equations~\eqref{eq:Diff},\eqref{eq:Poiss} and \eqref{eq:Laplace} describe the diffusion of point particles. 
To account for the volume of the diffusing molecule its radius is added to the excluded volume for the center of mass, see Fig~\ref{fig:Cartesian}(a). 
We depict the crowders as circles with radius $R$, but it is important to mention that any shape is possible for the crowding molecules.
The shape of the small (as compared to crowders) diffusing molecule is, however, restricted to circles or spheres with radius $r$.

\begin{figure}[H]
\centering
\subfigure[]{\includegraphics[width =.31\textwidth]{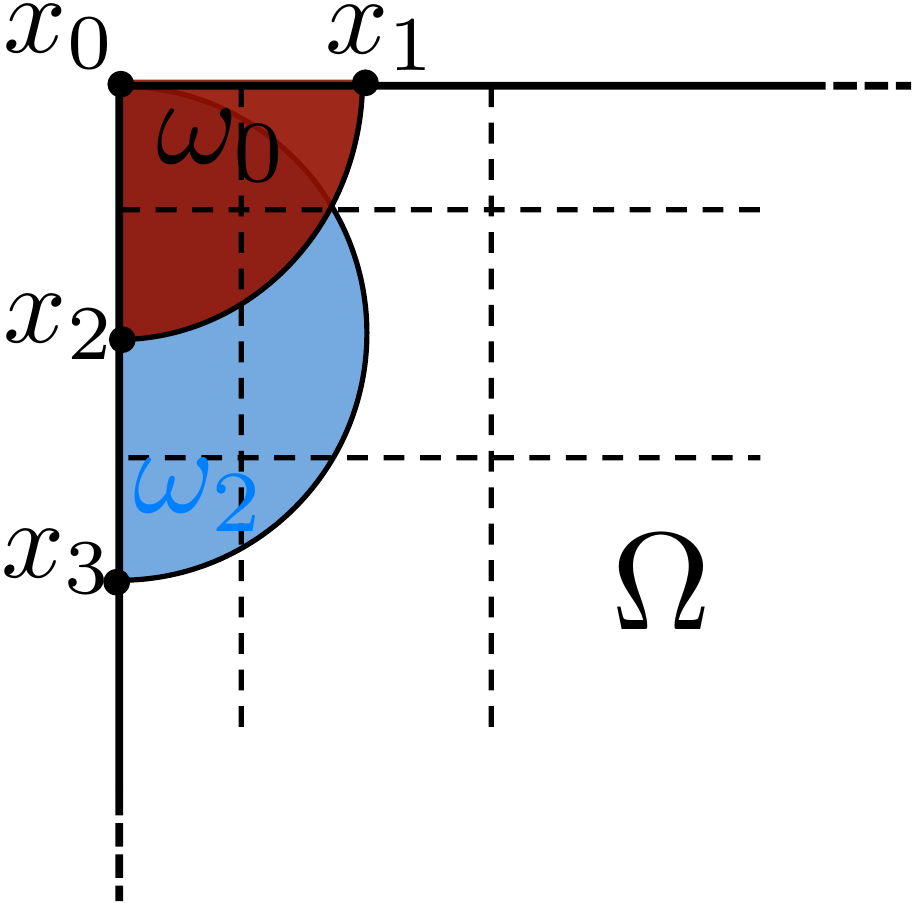} }
\subfigure[]{\includegraphics[width =.31\textwidth]{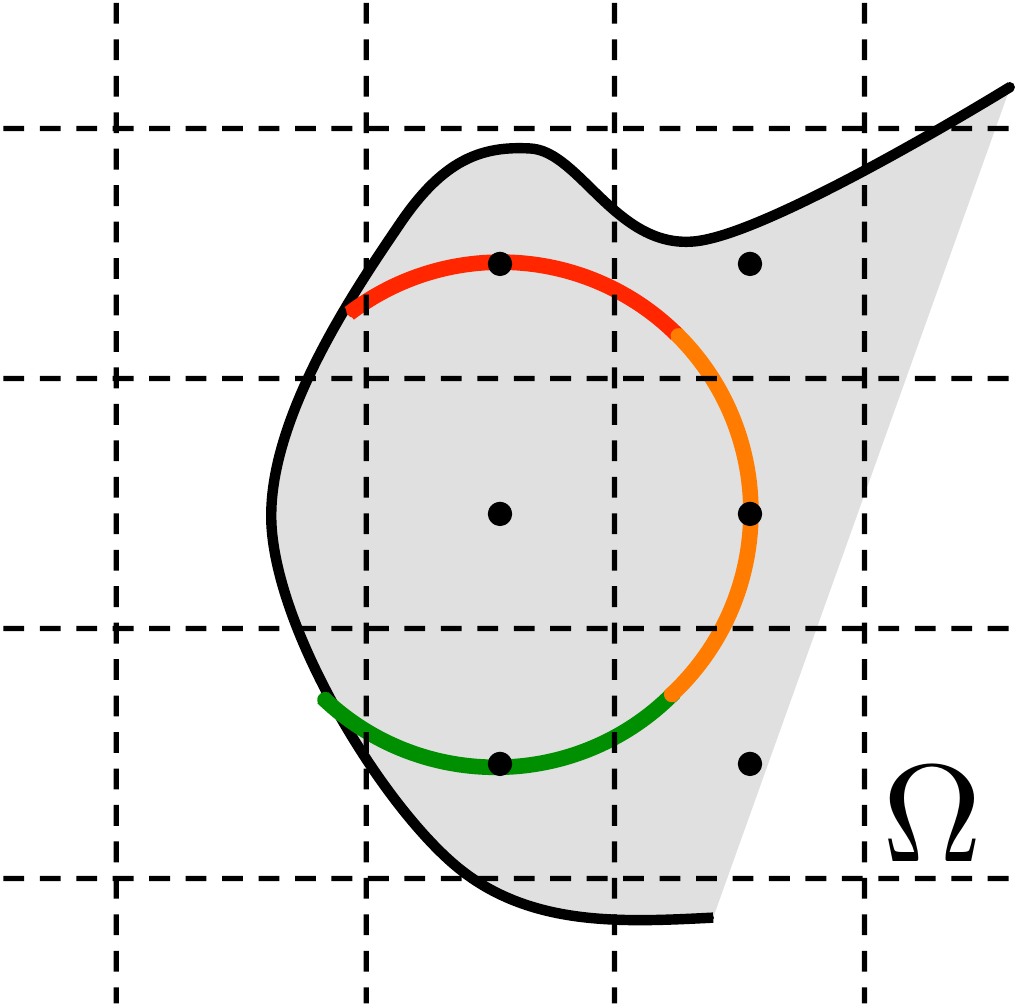} }
\subfigure[]{\includegraphics[width =.31\textwidth]{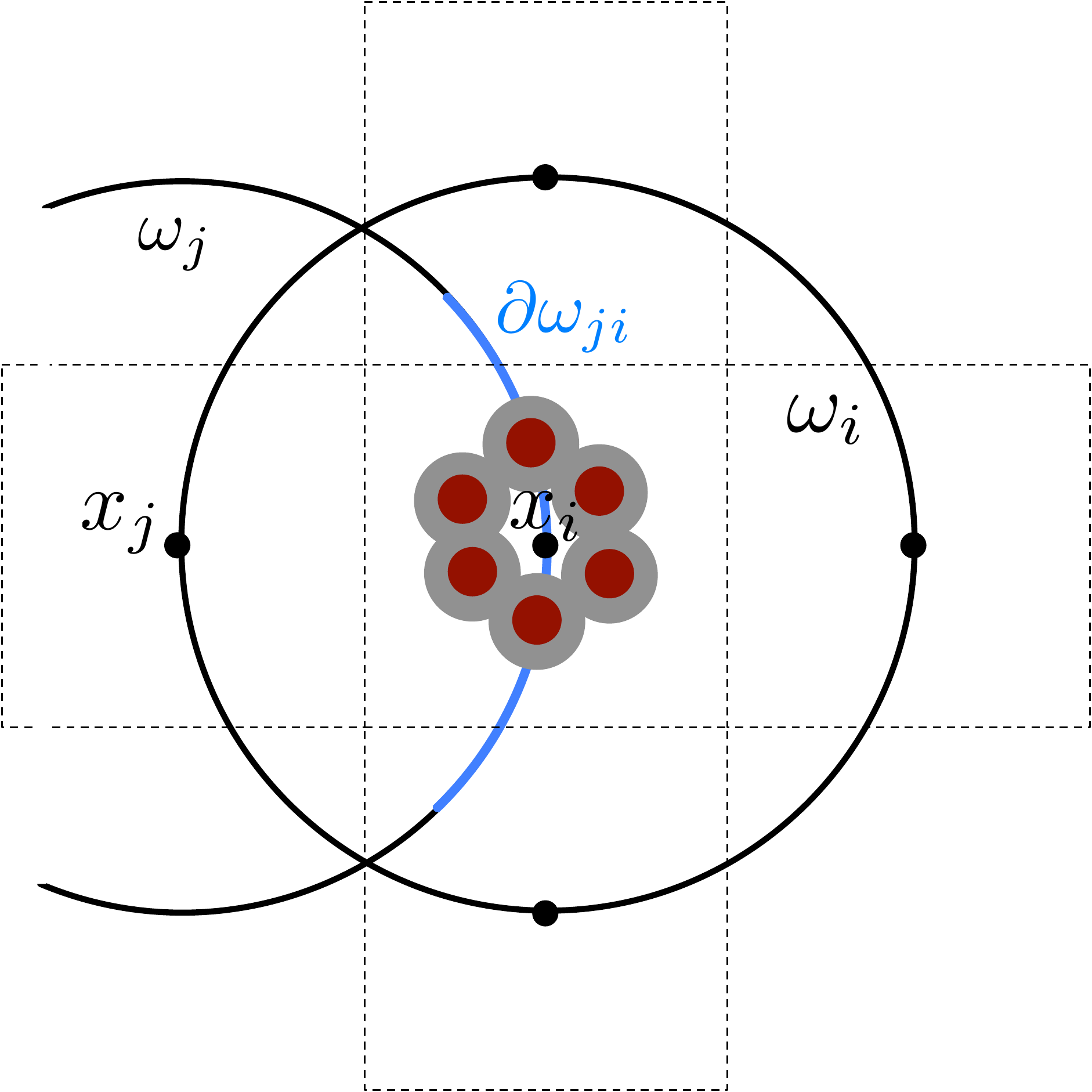} }
\caption{(a) Boundary treatment. (b) Using the first exit time to compute Cartesian jump rates for a curved domain $\Omega$. (c) Model error when interpreting molecules as well mixed and jumping between nodes.}
\label{fig:ModelProblem}
\end{figure}

Solving \eqref{eq:Poiss} and \eqref{eq:Laplace} numerically on the perforated domain $\omega_i$ means that the crowding molecules have to be resolved by a fine mesh. 
But, we have divided the global problem into $N$ local subproblems (one protective domain $\omega_i$ around each node $i$), which is only solved once and can be parallelized. 
This is similar to the approach in \cite{Brown2014} where deterministic local equations are solved on media with porous microstructures.
The stochastic simulation of the spatial SSA is then performed on the coarse mesh with $N$ nodes no longer resolving the individual obstacles.
The boundary conditions on the global domain $\Omega$ (reflective or absorbing) are implemented by posing these conditions on the secants of the half or quarter circles, which are the protective domains for boundary nodes, see Fig.~\ref{fig:ModelProblem}(a).
In Fig.~\ref{fig:ModelProblem}(b) we briefly illustrate how the first exit time approach can be further used to compute the jump rates for a Cartesian grid, discretizing a domain $\Omega$ with a curved boundary.
Here the part of $\partial\omega_i$ originating from the circle is imposed with the boundary conditions in \eqref{eq:Poiss} and \eqref{eq:Laplace} and the part originating from $\partial\Omega$ with the boundary condition valid on $\Omega$, which usually are reflecting or (partially) absorbing.
The circle is then divided in the same way as before to compute the splitting probabilities to the neighboring nodes remaining inside $\Omega$.

The simultaneous interpretation of the moving molecules being well-mixed inside the voxels $\calV_i$ and jumping from node $\fatx_i$ to $\fatx_j$ leads to problems when including crowders. 
Consider the case where just the center $\fatx_i$ is blocked, but voxel $\calV_i$ is sufficiently empty to be traversed, see Fig.~\ref{fig:ModelProblem}(c).
In this case the jump into voxel $\calV_i$, is possible ($\lambda_{ji}>0$), but the expected time to leave $\calV_i$ is infinity and hence the molecules get trapped inside $\calV_i$.
This does not agree with the microscopic situation, where molecules diffuse around $\fatx_i$.
To avoid unrealistic trapping, all jump propensities $\lambda_{ji}$ to voxels whose vertex is isolated are set to zero.
Equations \eqref{eq:Poiss} and \eqref{eq:Laplace} cannot be evaluated in $\fatx_i$ if the node is covered by the excluded volume and setting all $\lambda_{ji}$  to zero would over estimate the crowding effect considerably so we distribute the crowders such that $\fatx_i$ remains inside $\omega_i$.

\begin{figure}[H]
\centering
\subfigure[]{\includegraphics[width=0.32\textwidth]{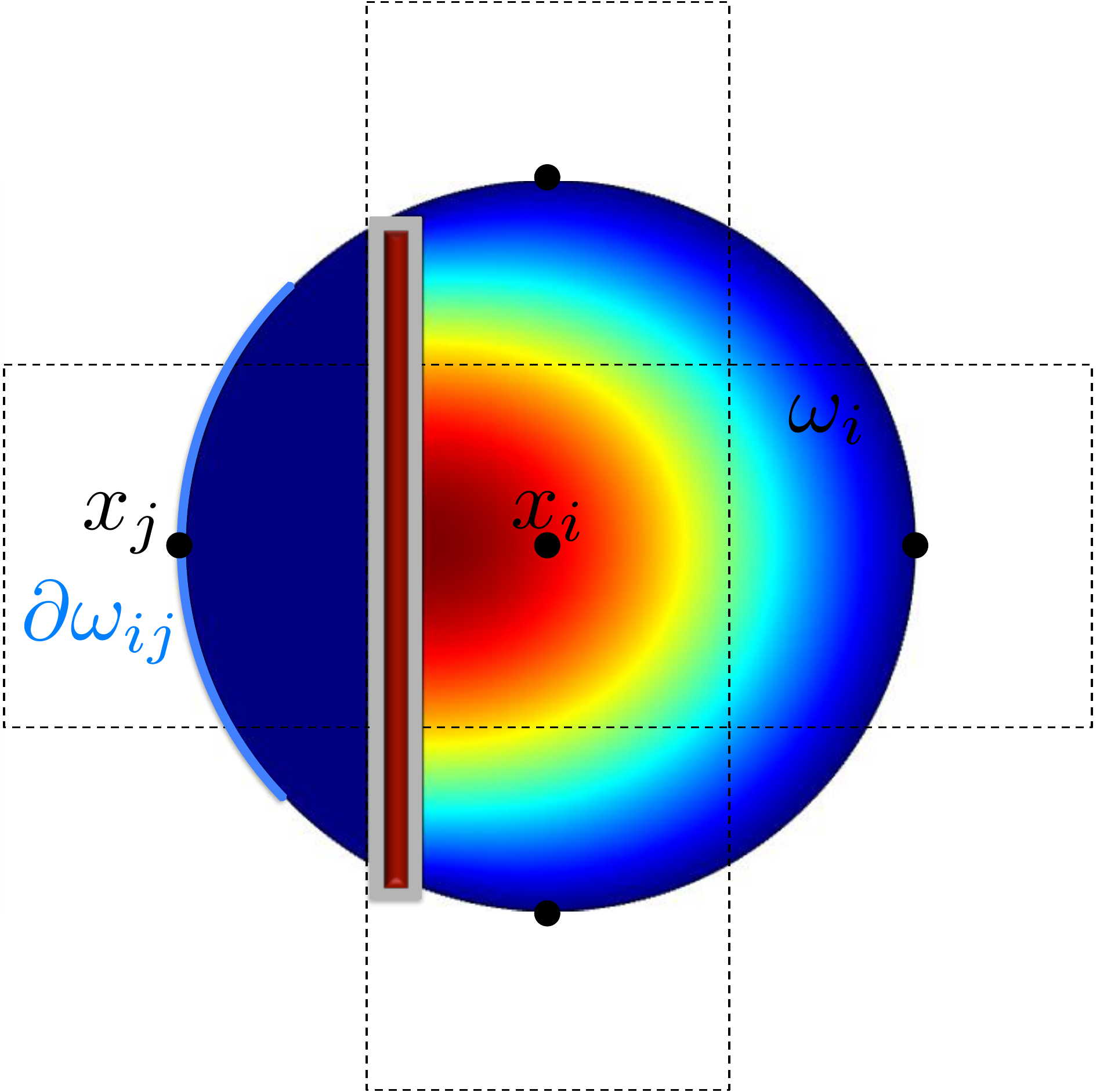}}
\subfigure[]{\includegraphics[width=0.32\textwidth]{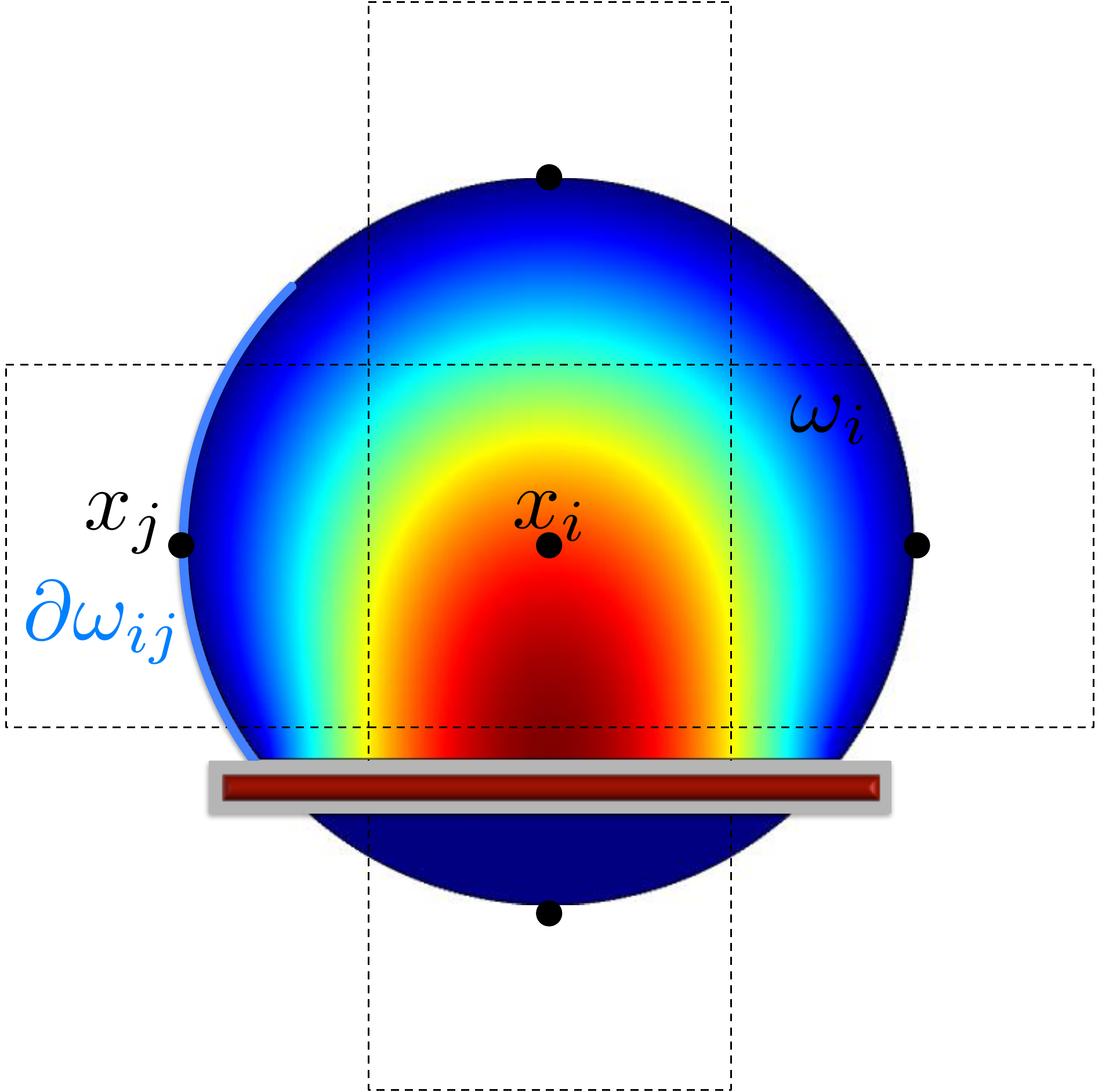}}
\caption{The effect of the microscopic position and orientation of the crowding molecules. (a) $\theta_{ij}=0$. (b) $\theta_{ij}> 0.25$.}
\label{fig:Positions}
\end{figure}

By using the expected exit time from $\omega_i$ around $\fatx_i$ to calculate the jump coefficients in a crowded environment it is the crowder distribution inside the whole circle $\omega_i$ that affects the coefficients $\lambda_{ii}$ and $\theta_{ij}$. 
This differs from other approaches to simulate diffusion in a crowded environment with a discrete space jump process. 
In \cite{Landman2011,Roberts2013,Taylor2015} it is only the percentage of occupied volume in the target voxel $\calV_j$ and in \cite{Grima2007} the difference in occupancy between $\calV_j$ and the origin $\calV_i$ 
that affect the jump rate $\lambda_{ij}=\theta_{ij}\lambda_{ii}$.
In our approach the microscopic positions of all crowding molecules inside $\omega_i$ are resolved and influence the jump coefficients.
In the case of non-spherical crowding molecules also the orientation is taken into account and long thin molecules with small volume can have a significant effect on $\lambda_{ij}$ and $\theta_{ij}$, see Fig.~\ref{fig:Positions}.
Note that in contrast to normal diffusion the jump propensities are no longer symmetric, i.e. in general $\lambda_{ij}\neq\lambda_{ji}$ and $\lambda_{ij}\neq\lambda_{im}$ for $j\neq m$.


\subsection{Statistics on the mesoscopic level}\label{sec:Stat}

Solving $N$ local problems where complicated geometries have to be resolved, see Fig.~\ref{fig:Cartesian}(c) is computationally expensive and will be inefficient if the crowding molecules move and the coefficients $\lambda_{ii}$ and $\theta_{ij}$ have to be recomputed often. 
Since the crowders' exact location is generally unknown we can compute statistics on a reference domain $\omega_i$ for given a percentage of occupied volume, a given shape and size of the molecules and $h$. 
Instead of solving $N$ PDEs of type \eqref{eq:Poiss} and $3N$ of type \eqref{eq:Laplace} at each time step, we can then sample the coefficients $\lambda_{ii}$ and $\theta_{ij}$ from these precomputed distributions. 
This will be especially applicable for moving crowders, where new coefficients can be drawn from the distributions on the time scale of their diffusion.



\section{Mesoscopic to macroscopic: a space dependent diffusion map}\label{sec:macro}
In this section we derive a macroscopic diffusion equation with a space dependent diffusion coefficient $\gamma(\fatx)$, representing the effect of macromolecular crowding. 
We approximate the mesosocpic jump process by Fickian diffusion with a constant diffusion coefficient $\gamma_i$ inside each voxel $\calV_i$.
The mesoscopic expected exit time $E_i$ from a node $\fatx_i$ or voxel $\calV_i$ is connected via ~\eqref{eq:EMeso} to this diffusion coefficient $\gamma_i$. 
So we obtain a modfied version of the macroscopic deterministic diffusion equation \eqref{eq:DiffFree}
\begin{equation}\label{eq:DiffGamma}
u_t(\fatx,t) = \nabla\cdot(\gamma(\fatx)\nabla u(\fatx,t)),\quad \fatx\in\Omega,
\end{equation}
where
\begin{equation}\label{eq:Gammax}
\left.\gamma(\fatx)\right|_{\calV_i} =\gamma_i= \frac{h^2}{2dE_i}. 
\end{equation}
For transferring the mesoscopic jump rate to the less detailed macroscopic level we only use $\lambda_{ii}$ and the random walk becomes symmetric in each voxel, i.e. $\theta_{ij}=\theta_{im}=0.25$ for $j\neq m$, but the asymmetry between back and forth jumps is preserved, i.e. $\lambda_{ij}\neq\lambda_{ji}$.
Alternatively $\gamma(\fatx)$ can be defined on the edges $\partial\calV_{ij}$ (see Fig.~\ref{fig:Cartesian}(b)) of voxel $\calV_j$ by $\gamma_{ij}=h^2/2(\lambda_{ij}+\lambda_{ji})$.
This corresponds to a mesoscopic jump process with symmetry in $\lambda_{ij}=\lambda_{ji}$ and non-symmetric jumps out of a box $\lambda_{ij}\neq\lambda_{im}$ for $j\neq m$.

The anomalous diffusion is here modeled by a space dependent diffusion coefficient.
In \cite{Blanc2015} molecules change their internal state, i.e. their diffusion constant, spontaneously in time.
This correlates to our model when the crowding macromolecules are moving and the diffusion constant hence also becomes time-dependent: $\gamma(\fatx,t)$.

In the next section we will use the diffusion map $\gamma(\fatx)$ to derive space dependent reaction rates.

%
%


\section{Reactions}\label{sec:reactions}
Approaches to correct the reaction rates for crowding effects use either time dependent reaction rates $k(t)$ \cite{Berry2002,Schnell2004} or a static modification \cite{Ellis2001,Grima2010,Grima2007}.
Similarly to the latter we will use the space dependent diffusion coefficient $\gamma(\fatx)$ from the previous section to compute mesoscopic reaction rates $k_{i}$ inside each voxel $\calV_i$.
This static rate becomes time-dependent $k_i(t)$ if we model moving crowders.
According to \cite{Hellander2015} the dilute mesoscopic rate $k_0$ for bimolecular reactions in a three dimensional cube of volume $h^3$ is linked to the effective rate $k_{CK}$ by Collins and Kimball \cite{CoKi} by
\begin{equation}
k_{0} = k_{CK}/h^3,\quad\quad\quad k_{CK} = \frac{4\pi\sigma\gamma_0 k_r}{4\pi\sigma\gamma_0+k_r},
\label{eq:ColKim}
\end{equation}
where $k_r$ is the intrinsic reaction rate and $\sigma$ the sum of the two reaction radii. In 2D there is no equivalent formula but approximations are derived in \cite{Fange2010,Hellander2012}.

Assuming constant Fickian diffusion inside each voxel as in Sec.~\ref{sec:macro}, we can now compute mesoscopic reaction rates $k_i$ for each voxel by inserting the $\gamma_i$ from \eqref{eq:Gammax} into \eqref{eq:ColKim}
\begin{eqnarray}
k(\fatx) &=& \frac{1}{h^3}\frac{4\pi\sigma\gamma(\fatx) k_r}{4\pi\sigma\gamma(\fatx)+k_r}\quad\text{and}\quad \left.k(\fatx)\right|_{\calV_i} =k_i.
\label{eq:ReactionRate}
\end{eqnarray}
In our macroscopic framework to model crowding by a space dependent diffusion map, these reaction rates $k_i$ model reactions under excluded volume effects and can be used in a mesoscopic simulation, where reactions inside each voxel $\calV_i$ have their specific reaction rate or in a macroscopic simulation, where a space dependent reaction term is included in the PDE \eqref{eq:DiffGamma}.

In this model only bimolecular associations are affected by macromolecular crowding, since the hindered diffusion changes the hitting time for the reaction partners.
In the internal states model \cite{Blanc2015} also birth-death processes and isomerizations become anomalous.
It is, however, questionable if it is meaningful to talk about birth-death processes, when considering excluded volume effects, since all reactants and products also occupy space.
With scaled particle theory \cite{Grasberger1986} also dissociation events are affected by the excluded volume which is due to two spherical molecules having a different activity coefficient than one molecule with the same total area.
In our model, dissociaton is not affected either since the two products are assumed to occupy the same area.

In Section~\ref{sec:Reactions}, we perform numerical experiments to examine for which parameters crowding molecules enhance or decrease the rate of biomolecular reactions.


\section{Numerical Experiments}

In the following experiments we solve \eqref{eq:Poiss} and \eqref{eq:Laplace} in 2D with COMSOL Multiphysics on $\omega_i,\,i=1\dots N$. 
To be able to evaluate the solutions at $\fatx_i$ the crowders are randomly distributed such that the nodes $\fatx_i$ remain inside the perforated domain $\omega_i$ and are not cut out by the excluded volume. 

\subsection{Effect of crowding on jump propensities}

%

\begin{figure}[H]
\centering
\subfigure[$r = 0$]{\includegraphics[width=0.31\textwidth]{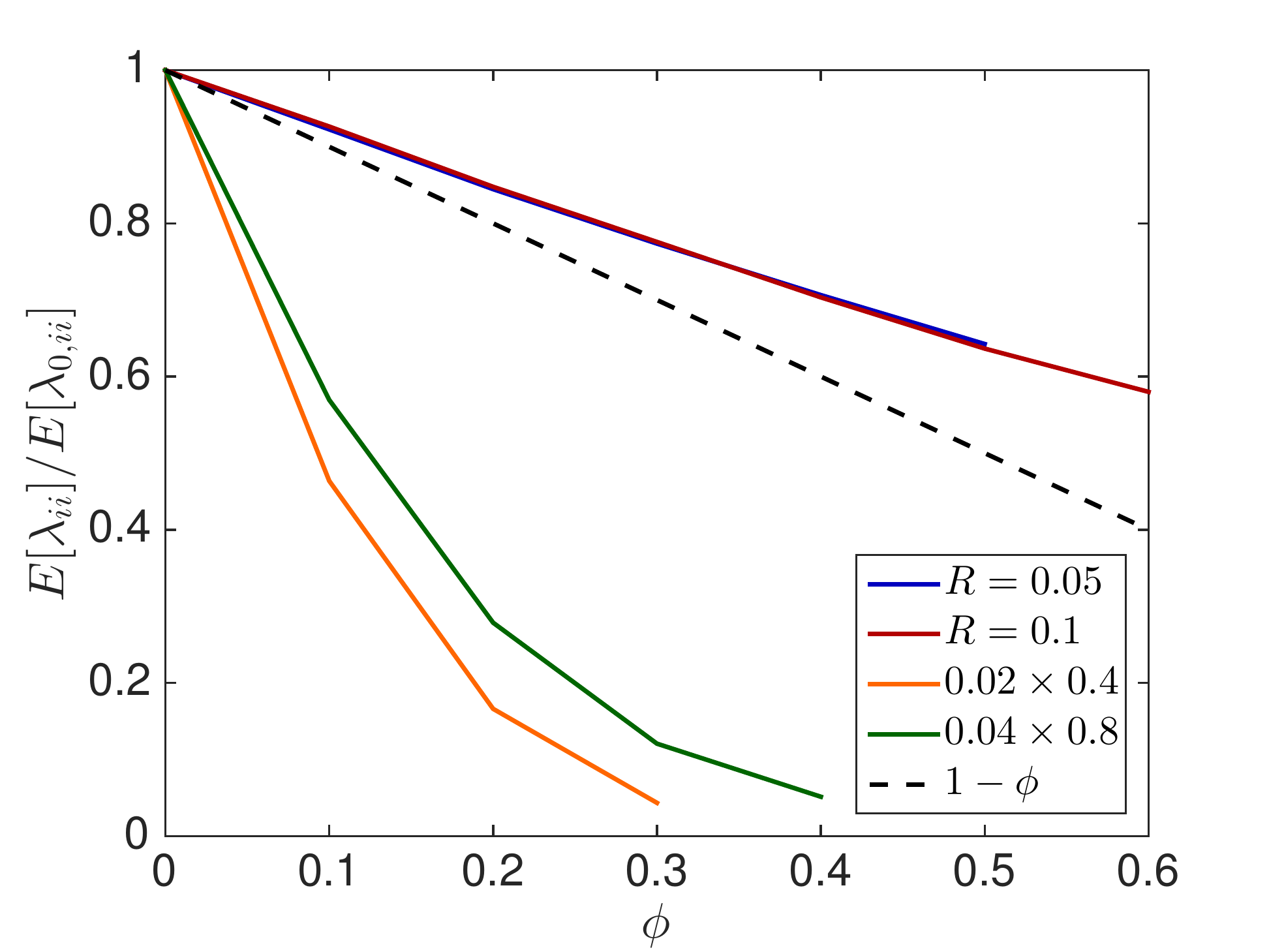}}
\subfigure[$r = 1e-4$]{\includegraphics[width=0.31\textwidth]{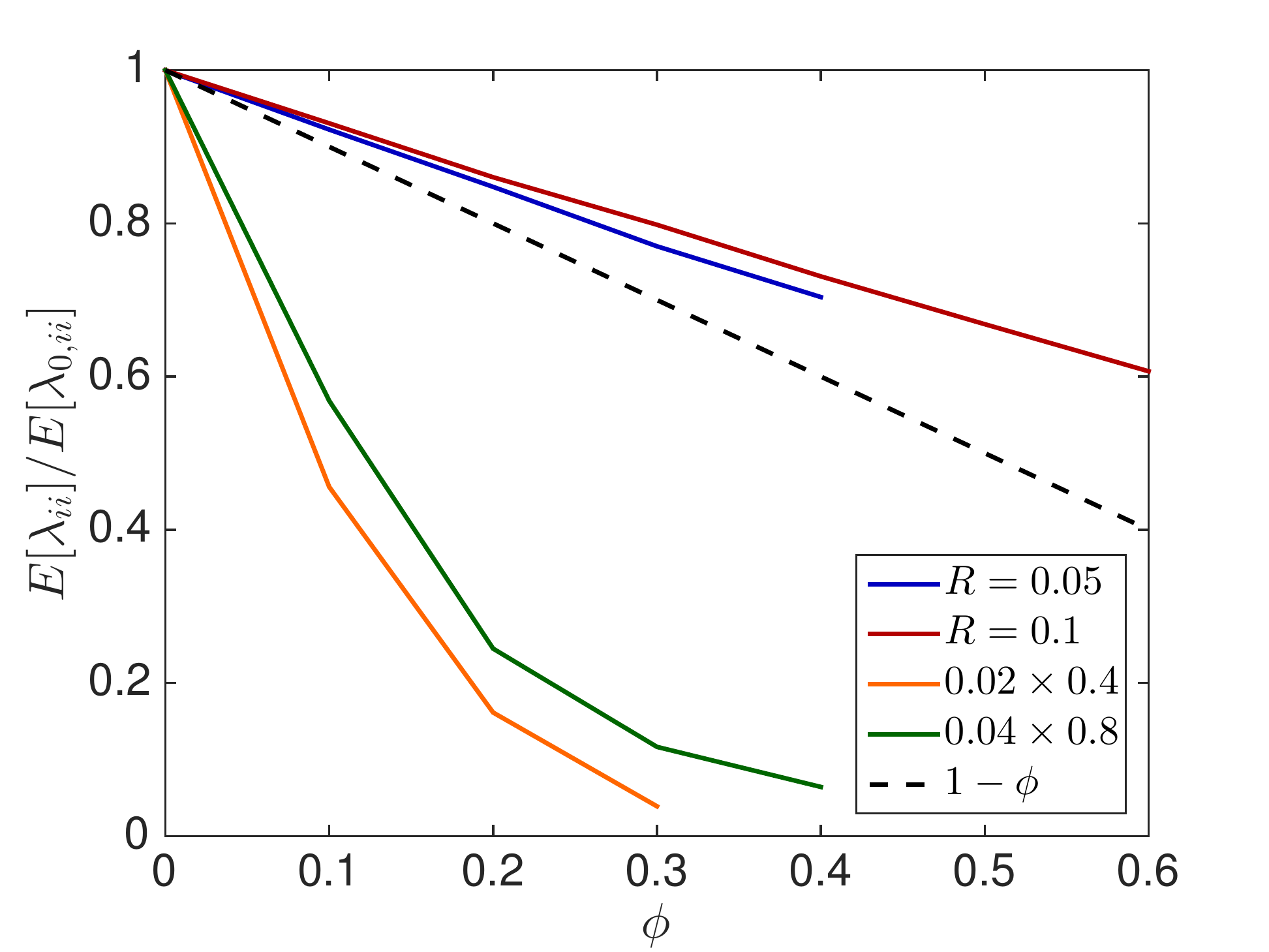}}
\subfigure[$r = 1e-2$]{\includegraphics[width=0.31\textwidth]{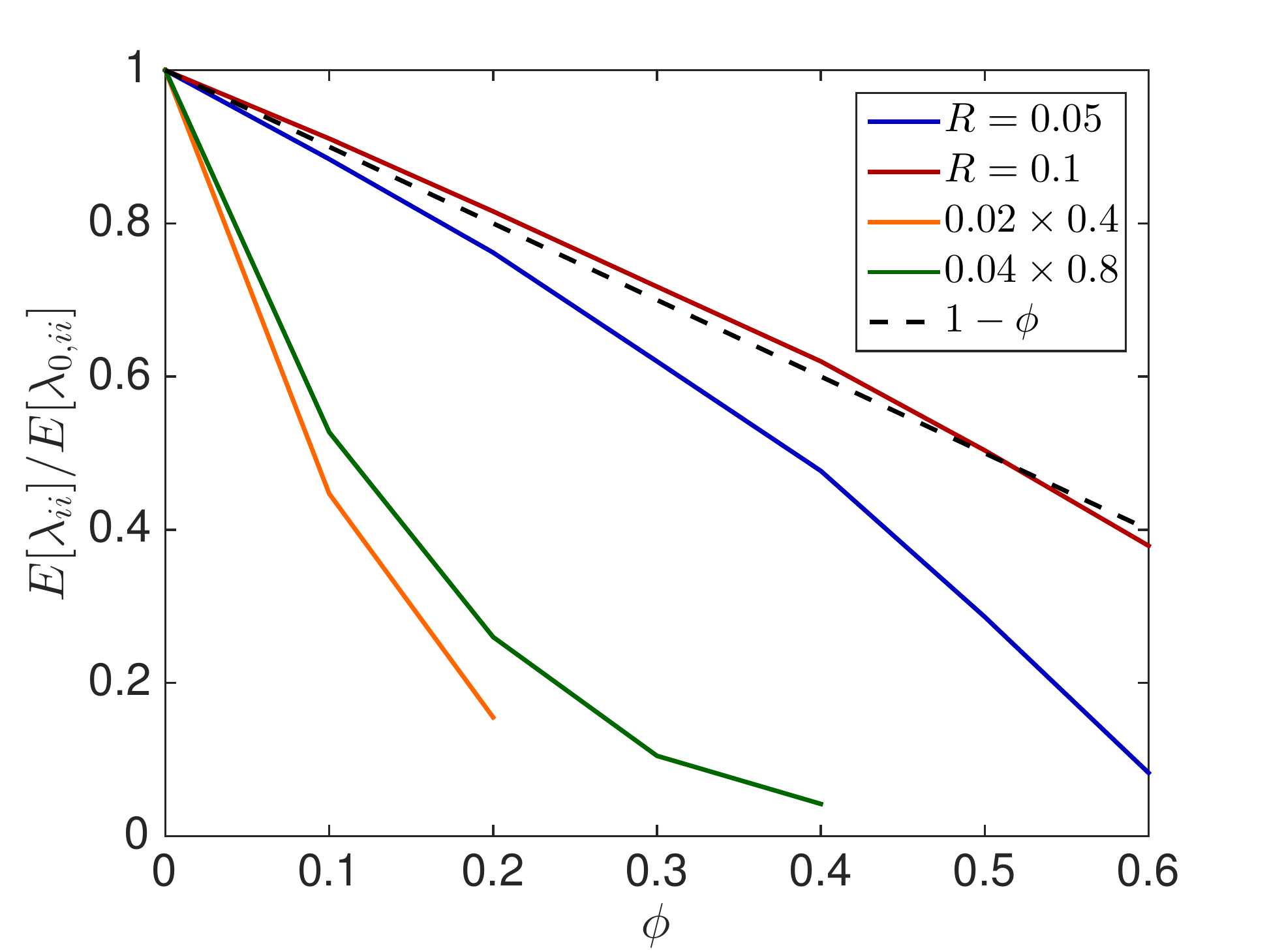}}\\
\subfigure[$r = 3e-2$]{\includegraphics[width=0.31\textwidth]{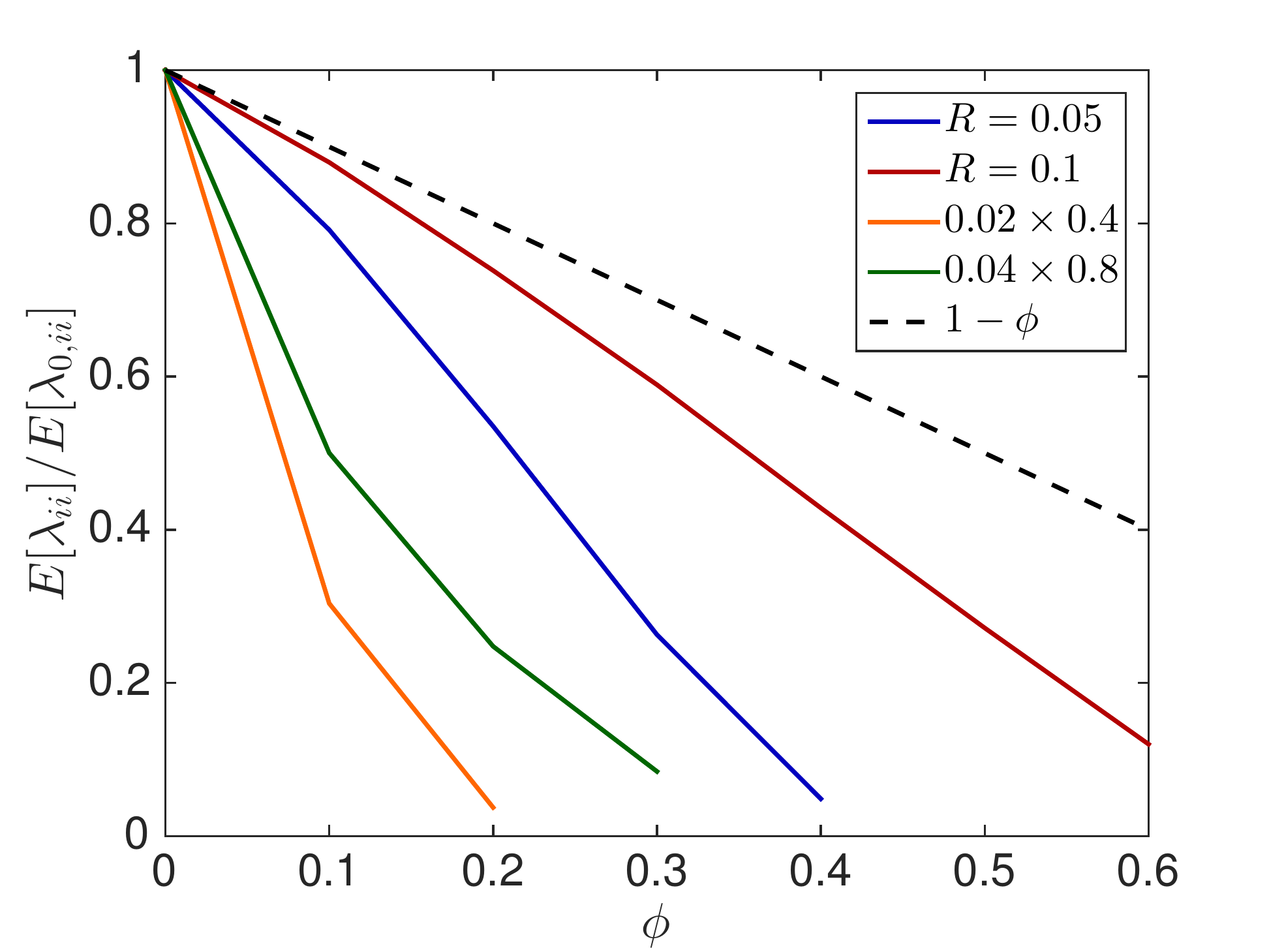}}
\subfigure[$r = 1e-1$]{\includegraphics[width=0.31\textwidth]{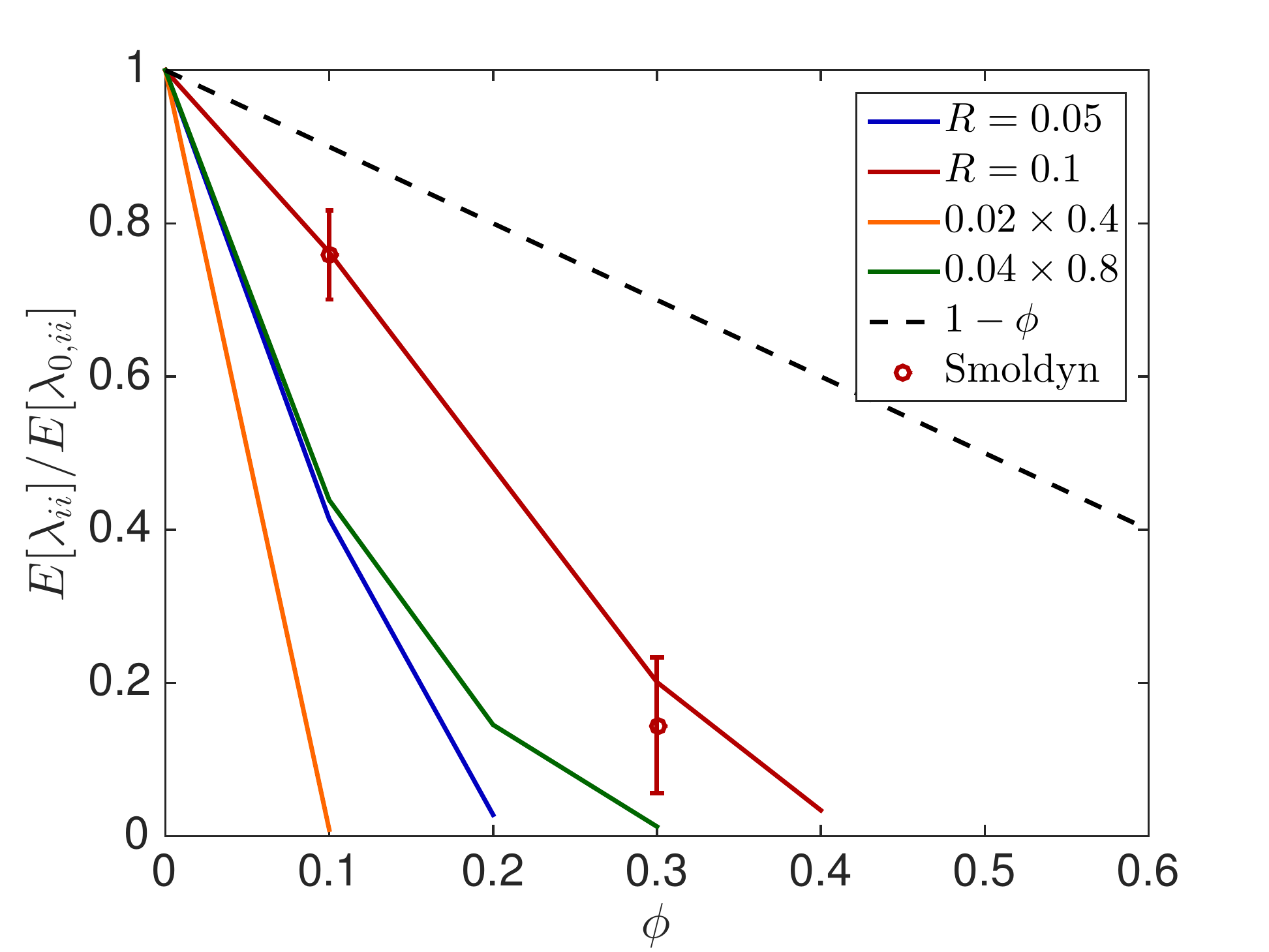}}
\subfigure[$r = 2e-1$]{\includegraphics[width=0.31\textwidth]{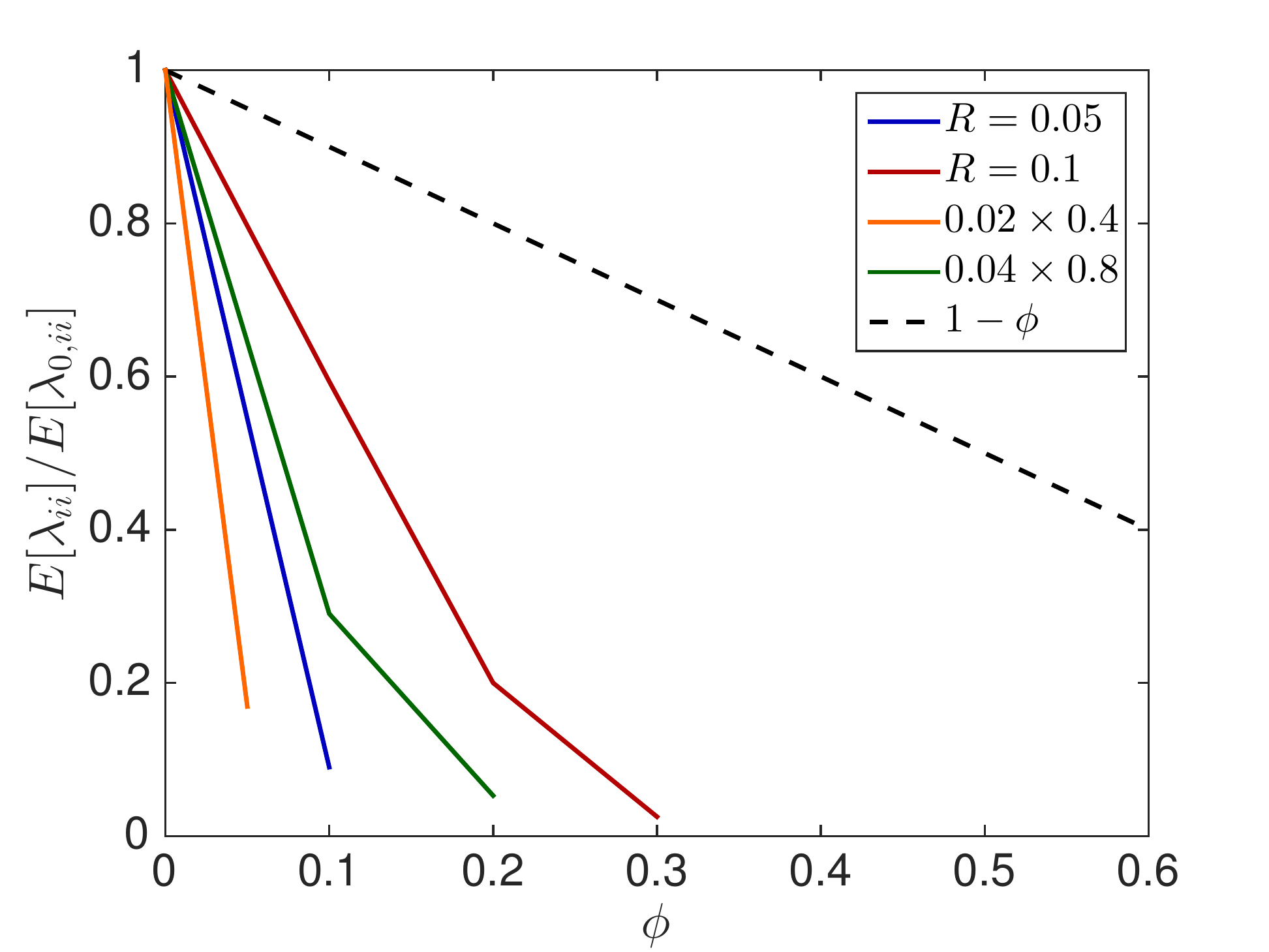}}
\caption{The mean value of the mesoscopic jump coefficients in the crowded environment $\mathbb{E}[\lambda_{ii}]$ compared to $\mathbb{E}[\lambda_{0,ii}]=4$ in dilute media. Averages are taken over $M = 100$ different crowder distributions. The obstacles are either small spheres (blue)/rectangles (orange) or larger spheres (red)/rectangles (green). The ratio of width to length is 20 for the rectangles. The spherical moving molecule has radius $r$. In (e) we compare the mesoscopic coefficients with the results from a Brownian dynamics simulaion, where we generate $1e4$ trajectories for $10$ different crowder distributions with the software Smoldyn.}
\label{fig:Stat}
\end{figure}

We first investigate how the jump propensity $\lambda_{ii}$ changes in different crowding situations.
We therefore compute $\lambda_{ii}$ on a reference domain $\omega_i$ with $h = 1$ and different crowders and sizes of the moving molecule $r$ and compare it to the jump rate $\lambda_{0,ij}$ in dilute medium.
In Fig.~\ref{fig:Stat}, we compare the mean value of the jump propensities $\mathbb{E}[\lambda_{ii}]$ for different distributions of crowders and an increasing percentage of occupied volume $\phi$ with the jump propensities when no crowders are present, where $\mathbb{E}[\lambda_{0,ii}]=\lambda_{0,ii}=4$. 
We test two different sizes of crowding molecules for both rectangles and spheres.
The reference line is the linear scaling where $\lambda_{ii}=(1-\phi)\lambda_{0,ii}$ as in \cite{Landman2011,Phillips2009}. 

\begin{figure}[H]
\centering
\subfigure[$r = 1e-4$]{\includegraphics[width=0.31\textwidth]{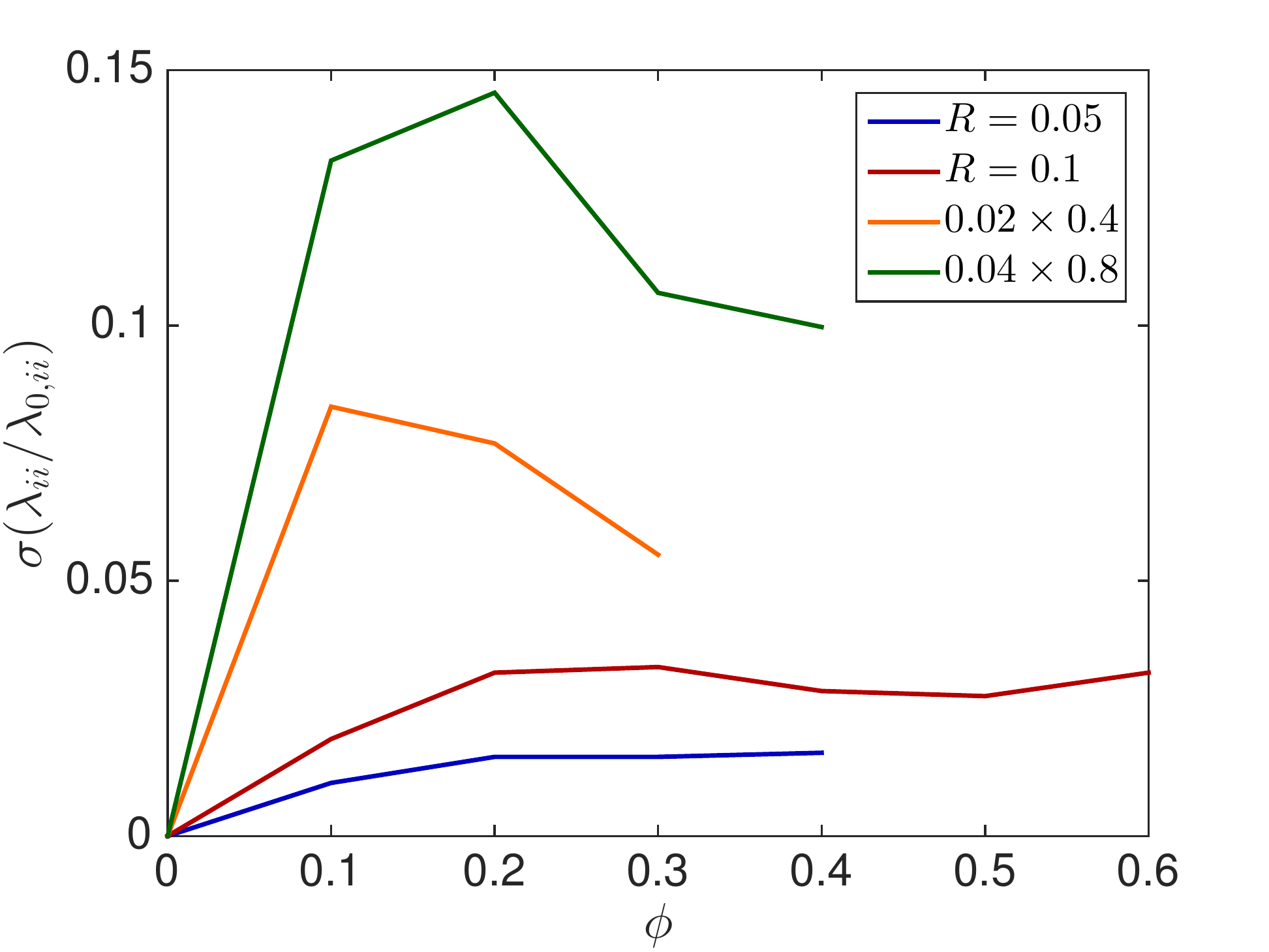}}
\subfigure[$r = 3e-2$]{\includegraphics[width=0.31\textwidth]{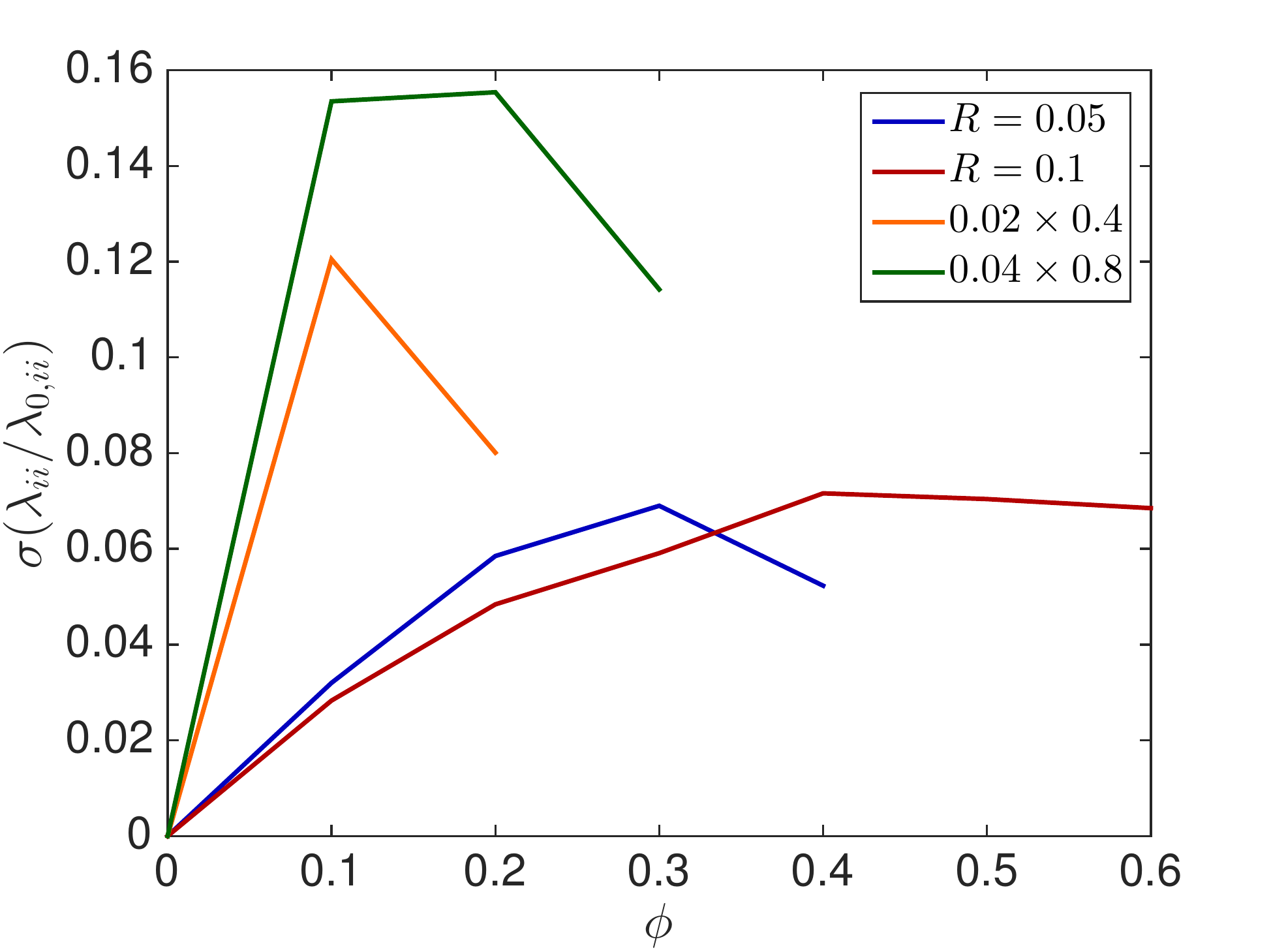}}
\caption{Standard deviation $\sigma(\lambda_{ii}/\lambda_{0,ii})$ for $M=100$ different crowder distributions for the obstacles in Fig.~\ref{fig:Stat}.}
\label{fig:Sigma}
\end{figure}

We observe that small obstacles (blue and orange in Fig.~\ref{fig:Stat}) hinder diffusion more than big crowders for the same percentage of occupied volume, since they have more reflecting surfaces than larger crowders.
The same holds for elongated rectangular crowders (green and orange in Fig.~\ref{fig:Stat}), as they create long barriers without occupying a lot of volume.
An increasing size $r$ of the diffusing molecule leads to an increase in the crowding effect, which is intuitive, since a bigger molecule finds less holes through which to escape.
These results agree with the findings in \cite{Muramatsu1988} and \cite{Ellery2015}.
We see that the averaged linear reduction of the jump propensity can be a good model when the diffusing molecule is about a tenth of the size of the crowders, but over- or underestimates the effect of occupied volume when the diffusing species is smaller or bigger, respectively.
Since an average protein has a radius of ca. 2nm \cite{Phillips2009} and the biggest macromolecules in the cell, the ribosomes, have a radius of up to 15nm the linear correction is a good approximation for many scenarios. 
The reduction in jump propensity, however, starts to behave exponentially, as in  \cite{Grima2007}, for large diffusing molecules.
The case $r=0$ corresponds to a point particle which is irrelevant when simulating excluded volume effects, but we include it to show the limit for very small particles.
To confirm these mesoscopic jump rates we compare them to the inverse of the expected exit time computed by a Brownian dynamics simulation.
We simulate $1e4$ trajectories with the open source software Smoldyn \cite{AnAdBrAr10,Andrews2004} for 10 different crowder distributions and equally sized crowding and moving molecules with $R=r=1e-1$ and see in Fig.~\ref{fig:Stat}(e) that the computationally expensive microscopic results agree well with the mesoscopic coefficients.

In Fig.~\ref{fig:Sigma}, the standard deviation $\sigma(\lambda_{ii}/\lambda_{0,ii})$ initially increases as more and more crowders are added but converges towards zero when the system approaches the state where no escape to the boundary is possible.

Simply rescaling the jump propensity $\lambda_{ii}$ for each node $i$ by a constant factor will lead to normal diffusion at reduced rate.
To observe anomalous diffusion in the crowded environment it is helpful to investigate the mean square displacement (MSD).

\subsection{The mean square displacement}
\begin{figure}[H]
\centering
\subfigure[]{\includegraphics[width=0.3\textwidth]{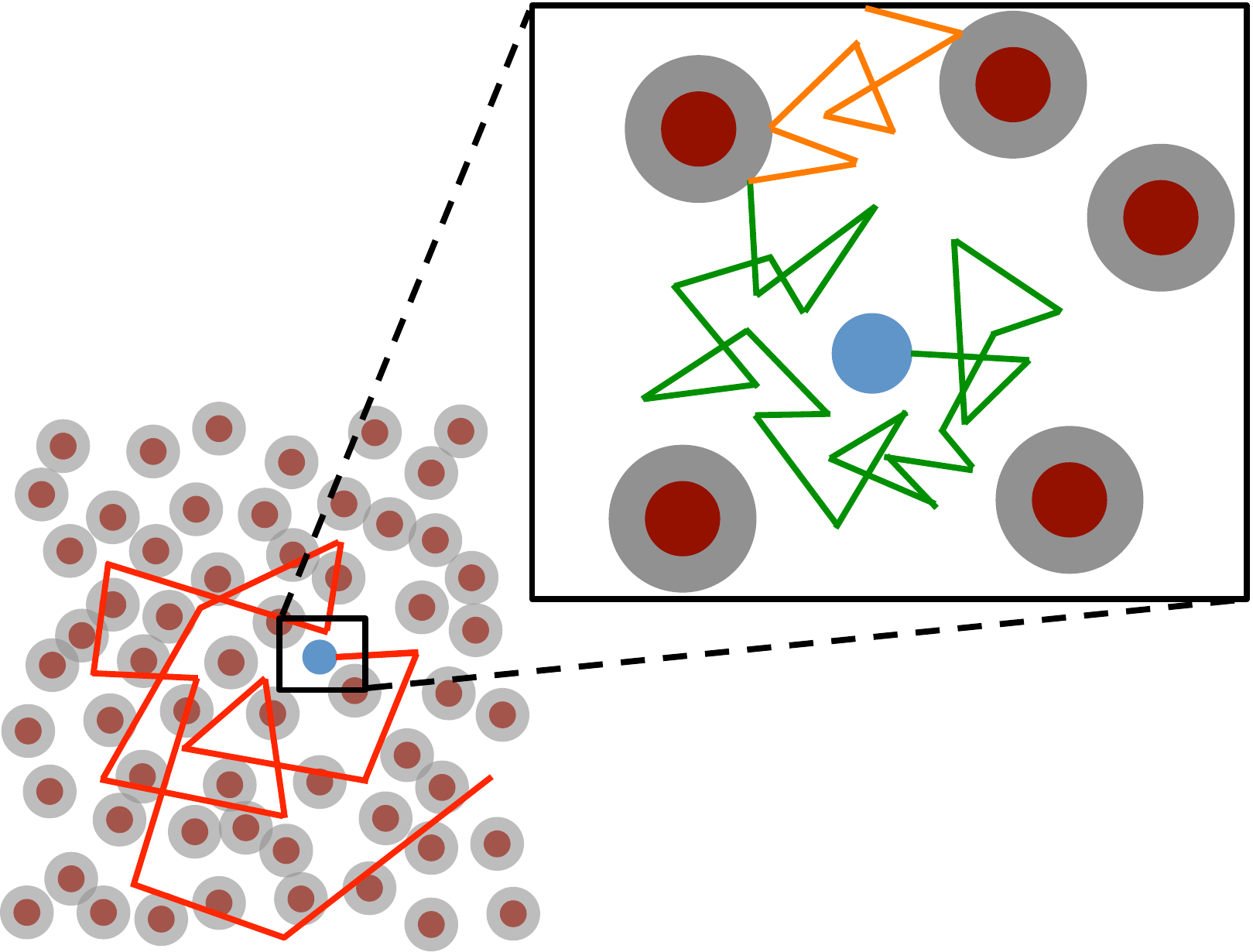}}
\hspace{1cm}
\subfigure[]{\includegraphics[width=0.45\textwidth]{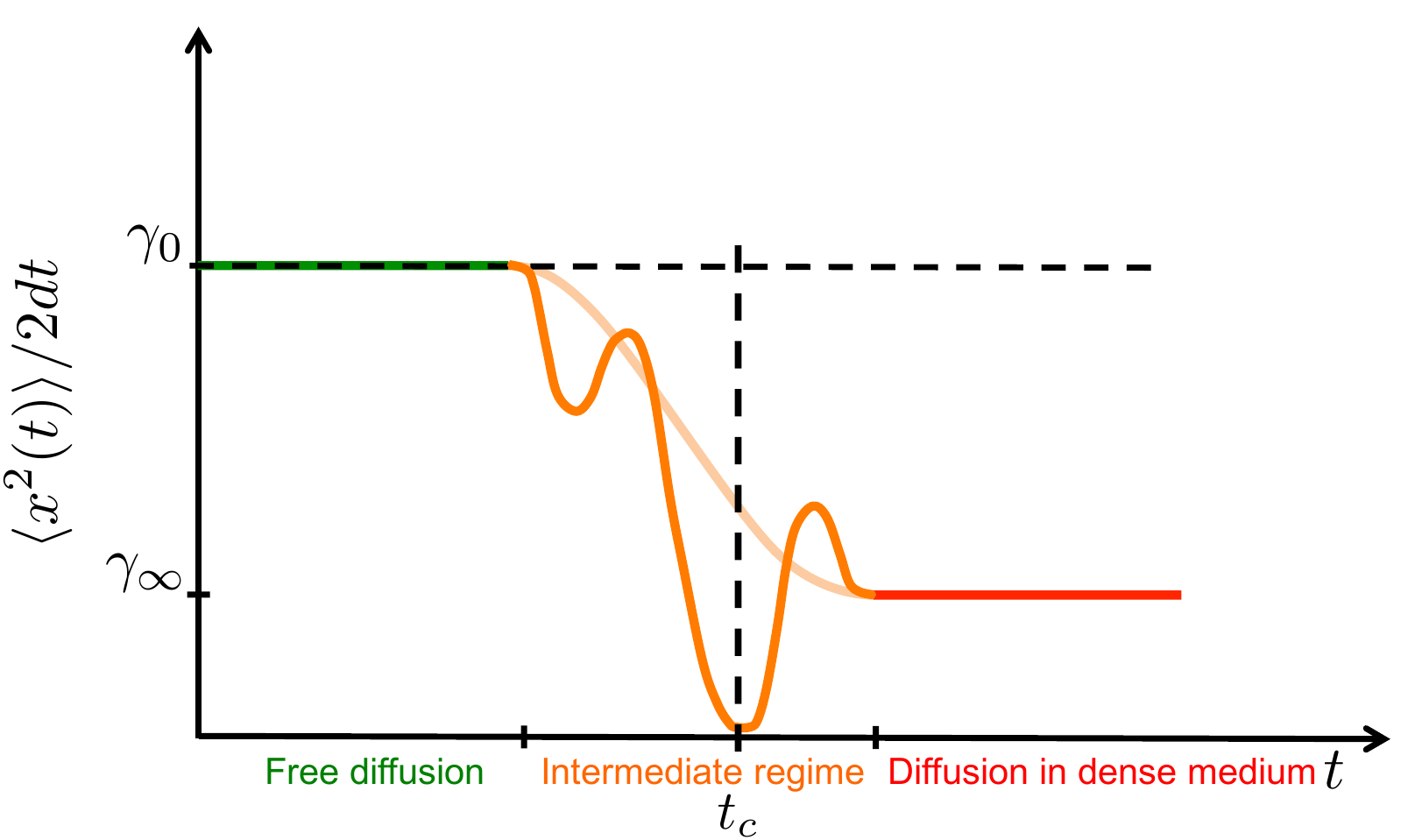}}
\caption{(a) Diffusion in the crowded cell environment: Initial free diffusion with $\gamma_0$ (green). After colliding with the first macromolecules the observed diffusion is slowed down and the molecules diffusion coefficient decays (orange). Long time behavior of slower diffusion with constant $\gamma_\infty$ in a dense medium. (b) MSD curve for diffusion in a crowded medium (solid line) and as reference normal diffusion (dashed line). The pale line corresponds to an ideal well mixed medium and the dark line to a realistic medium with stochastic fluctuations in the positions of the crowders.}
\label{fig:Crowding}
\end{figure}

As mentioned in Section \ref{sec:Crowding} the MSD is linear in time for normal diffusion
\begin{equation}
\langle \fatx^2(t)\rangle = 2d\gamma_0t,
\end{equation}
but for anomalous diffusion the relation is no longer linear
\begin{equation}
\langle \fatx^2(t)\rangle = 2d\gamma_0t^\alpha,
\end{equation}
where $\alpha<1$ for subdiffusion.
In \cite{DiRienzo2014,Mommer2009} it was shown that diffusion in a crowded environment can be modeled by a temporal change of the diffusion constant.
First the molecules diffuses normally with rate $\gamma_0$ for very short time-scales, before it undergoes a transient anomalous phase with a changing diffusion coefficient $\gamma$ and finally stagnates into normal diffusion at a lower diffusion rate $\gamma_\infty$.
The initial normal diffusion represents the time the molecule diffuses in the solution before it encounters the first adjacent macromolecule and is slowed down by collisions.
On a large time scale the molecule appears to diffuse in a denser medium instead of around obstacles, hence the reduced diffusion rate $\gamma_\infty$, see Fig.~\ref{fig:Crowding}.
If the crowding macromolecules are distributed evenly the MSD decays monotonically between $\gamma_0$ and $\gamma_\infty$ (pale line), but due to stochastic variations in the medium it fluctuates before converging to $\gamma_\infty$ (dark line).
In Fig.~\ref{fig:Crowding}(a) we only depict collisions with the macromolecules responsible for excluded volume effects, but note that there are many collisions with the much smaller solvent molecules responsible for the Brownian motion.

Since we choose $h> R$ a jump in the mesoscopic model spans a number of macromolecules and the initial free diffusion phase is not resolved and we start to observe diffusion after the first jump of length $h$, that occurs after a critical time $t_c$ which can be approximated by
\begin{equation}
t_c\sim\frac{h^2}{2d\gamma_0}.
\end{equation}

In the following we will plot the $\langle \fatx^2\rangle/2d\gamma_0t$ in log-log-scale for different crowding situations to examine when anomalous behavior occurs.
Let $\fatp(t)\in\mathbb{R}^N$ be the probability vector for a diffusing molecule, such that $p_i(t)$ is the probability that the molecule is in voxel $\calV_i$ at time $t$.
As described in Section~\ref{sec:Spatial}, $\fatp(t)$ evolves by the master equation
\begin{equation}
\fatp_t = \fatD\fatp(t), \quad \fatp(0) = \fatp_0,\label{eq:pODE}
\end{equation}
where $D_{ij}=\lambda_{ji}$ for $i\neq j$ and $D_{ii}=-\lambda_{ii}$. The initial probability distribution $\fatp_0$ is $(0,\dots,1,\dots,0)^T$ with one at the starting node $\fatx_0$.
We choose the discretization such that $\fatD$ is small enough to solve \eqref{eq:pODE} numerically and compute the mean square displacement by
\begin{equation}
\langle (\fatx(t)-\fatx_0)^2\rangle = \sum_{i=1}^Np_i(t)(\fatx_i-\fatx_0)^2.
\end{equation}
In the following experiments we discretize the square $[0,1]\times[0,1]$ into $N=n^2$ voxels with space discretization $h=1/(n-1)$.
If not mentioned otherwise we release the molecules in $[0.5,0.5]$ at time $t=0$ and choose $n=41$.
To avoid boundary effects we set homogeneous Dirichlet boundary conditions on $\partial\Omega$ and show the solutions as long as more than $99\%$ of the mass is preserved, i.e. $\sum_{i=1}^Np_i(t)>0.99$.

\subsubsection{The effect of $\gamma_0$ and $\phi$}
\begin{figure}[H]
\centering
\subfigure[]{\includegraphics[width=.4\textwidth]{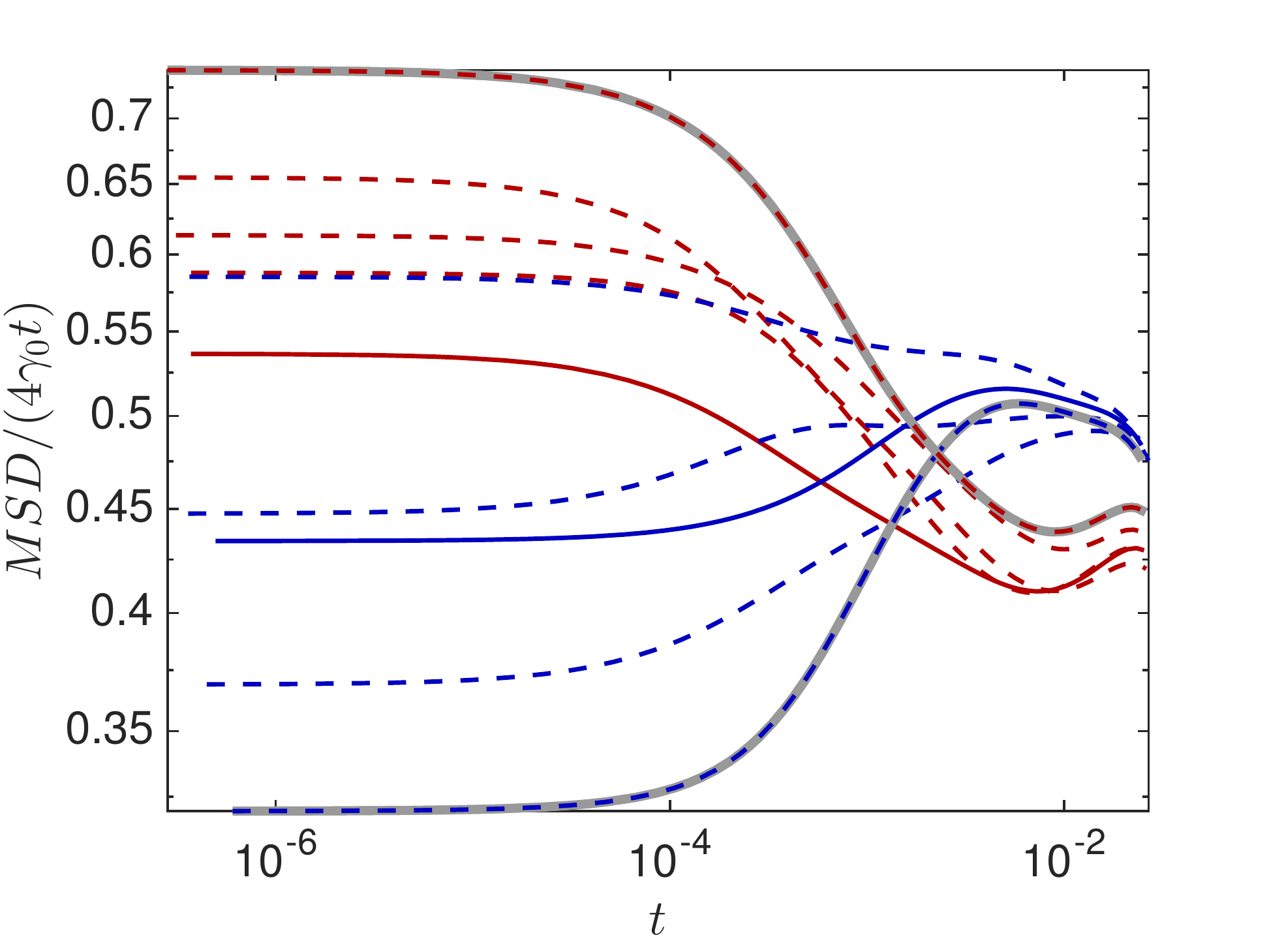} }
\subfigure[]{\includegraphics[width=.4\textwidth]{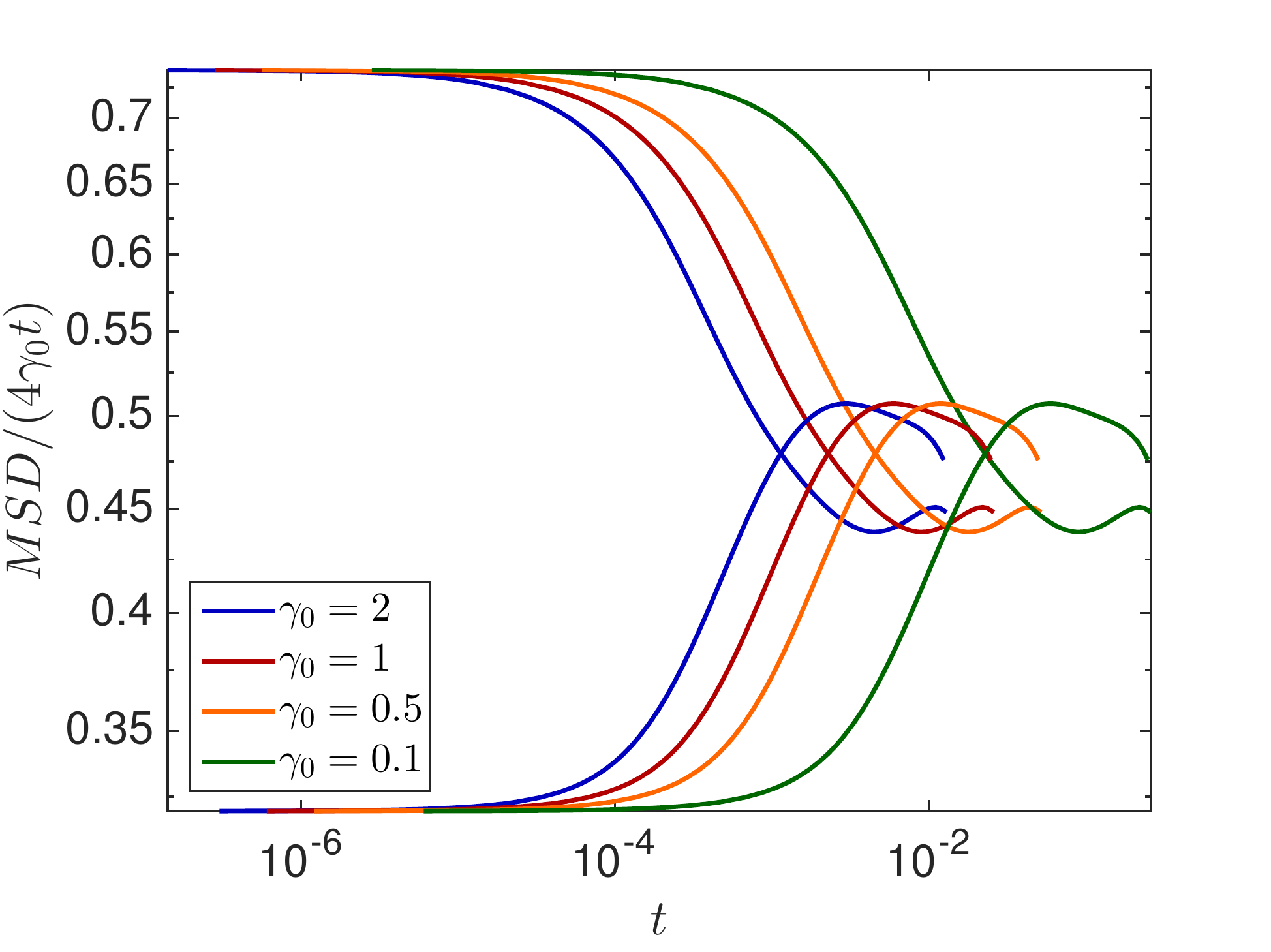} }\\
\subfigure[]{\includegraphics[width=.4\textwidth]{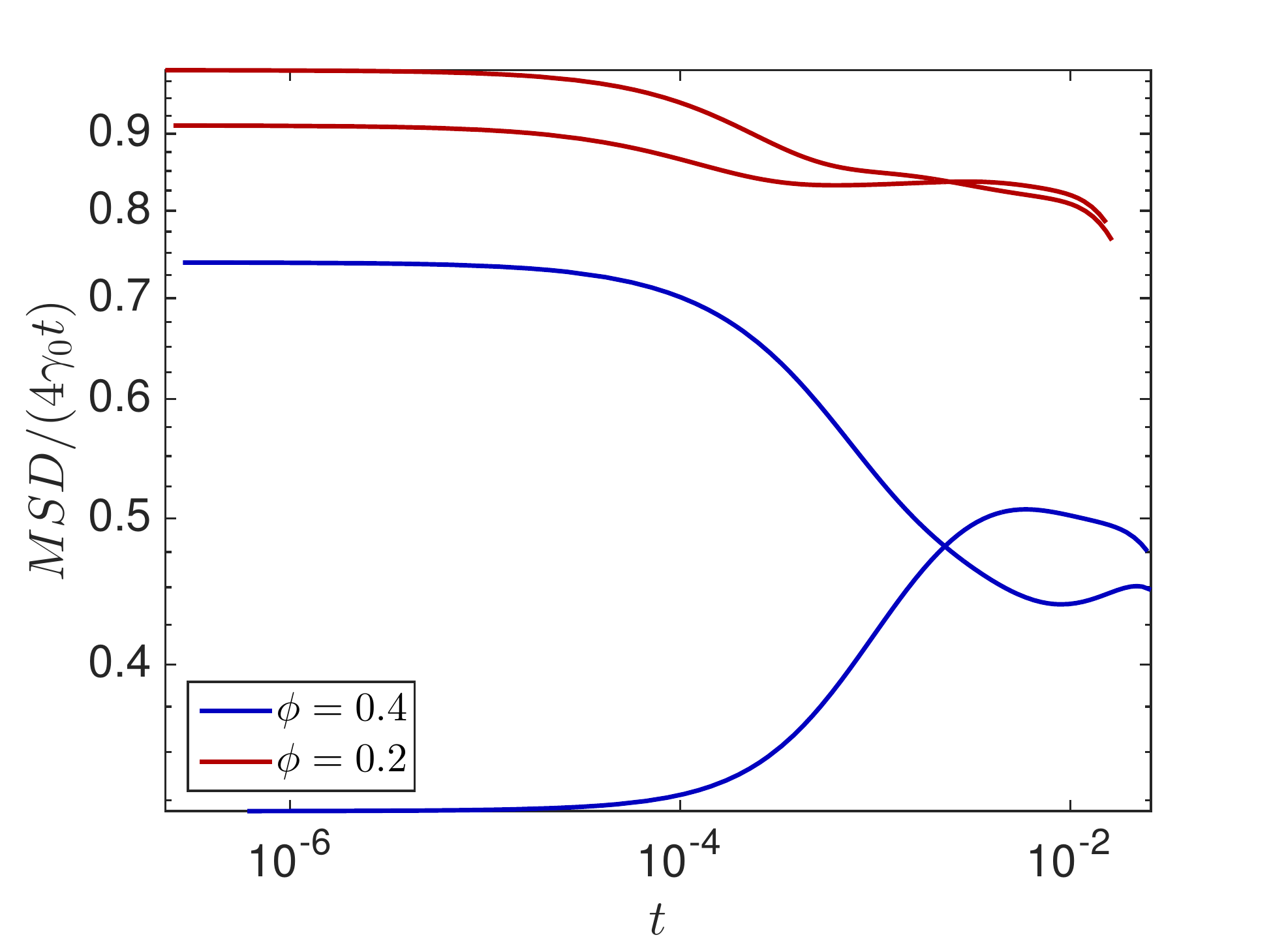} }
\subfigure[]{\includegraphics[width=.4\textwidth]{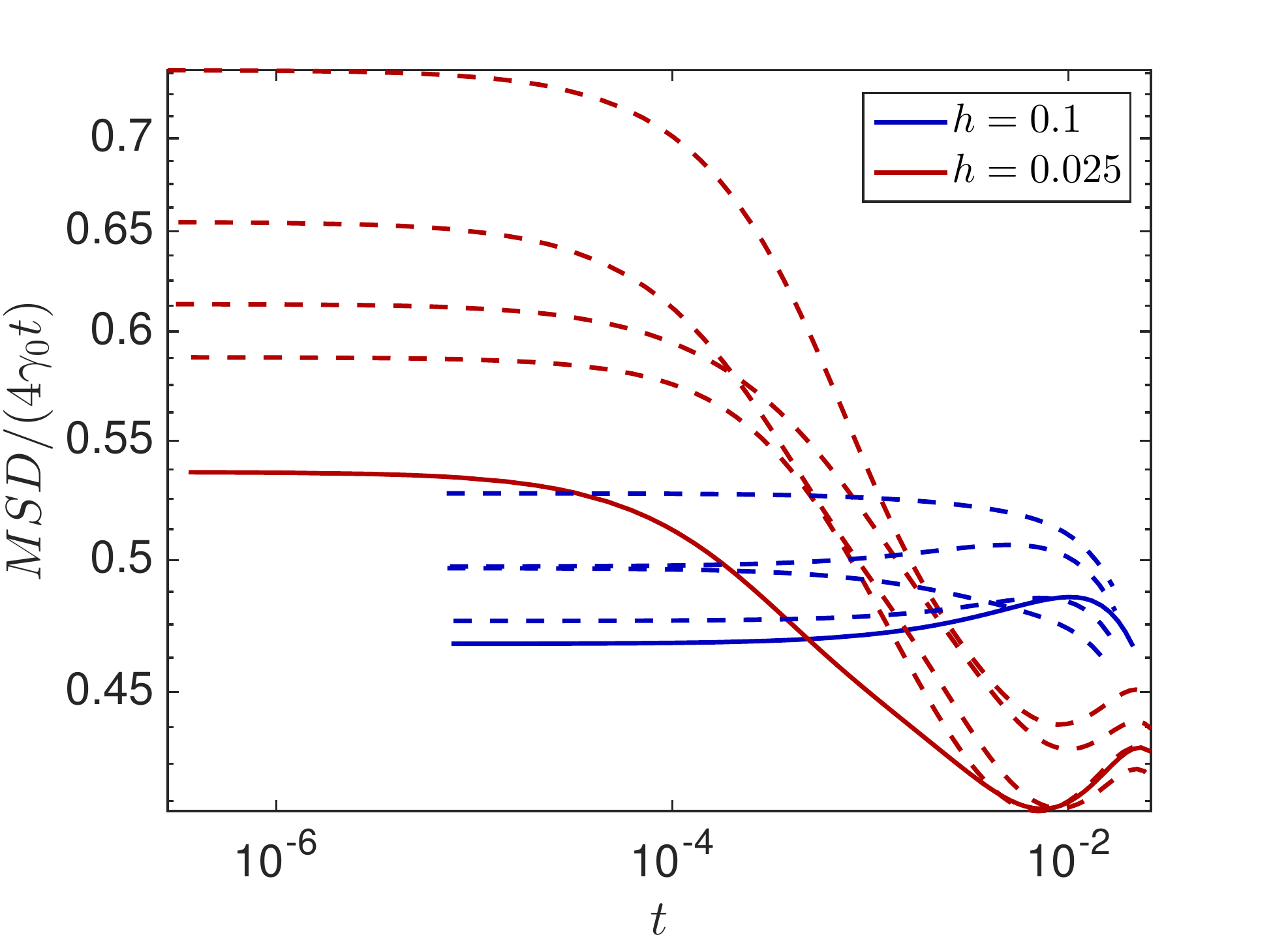} }
\caption{The MSD on a mesoscopic grid with $h = 0.025$ for different distributions of spherical crowders with $R=5e-3$ and a moving molecule with $r=5e-4$. (a) Different crowder distributions (red and blue) and different starting positions (dashed lines) for $\phi=0.4$ and $\gamma_0=1$. (b) Different $\gamma_0$ for the distributions highlighted in grey in (a). (c) Different $\phi$ for $\gamma_0 = 1$. (d) Different $h$ for the red distribution in (a) and $\fatx_0=[0.5,0.5]$.} 
\label{fig:MSD}
\end{figure}

In Fig.~\ref{fig:MSD}(a) we plot $MSD/(4\gamma_0t)$ in the crowded environment for different distributions of crowders.
Lines in the same color show diffusion in the same environment with starting positions $[0.5,0.5]$ (solid) and $[0.5\pm h,0.5\pm h]$ (dashed).
We clearly observe the anomalous behavior since $MSD/(4\gamma_0t)$ is not constant in time and the fluctuations due to the variations of the local environment around the starting position, but for longer times they converge towards the same long time behavior, before the boundary effects become apparent.

We choose the distributions and starting positions of the curves highlighted in grey to examine the effect of $\gamma_0$ and $\phi$ on the MSD.
In Fig.~\ref{fig:MSD}(b) we observe that the diffusion constant $\gamma_0$ only affects when the molecule undergoes anomalous diffusion but the length of the anomalous phase and the long time behavior are independent of $\gamma_0$.
The percentage of occupied volume $\phi$ on the other hand changes both, the average diffusion constant in the long time behavior and the duration of the transient regime of anomalous diffusion, as expected.

\subsubsection{Dependence on the space discretization $h$}
The mesoscopic model is designed for a voxel size $h$ considerably larger than the molecular radius in order to save computational effort compared to a microscopic simulation.
For $h\to 0$ the dilute and well-mixed assumptions in each voxel do no longer hold and the mesoscopic model is known to break down for the simulation of bimolecular reactions \cite{Isaacson2009}.
Different corrections to the reaction rates have been suggested \cite{Gillespie2013} and the references therein, but a minimal $h_{min}>R$ remains and space cannot be resolved any finer in the mesoscopic model.
For a finer resolution one has to switch to microscopic models, such as BD or CA
and we examine the effect of $h$ only for $h\gg R$.
A larger $h$ shifts the critical time $t_c$ after which we start to observe the molecule's motion to the right in Fig.~\ref{fig:Crowding}(b) , so for very large $h$ we will only see the long time behavior.
In Fig.~\ref{fig:MSD}(d) we see that the initial faster diffusion with $\gamma\sim\gamma_0$ is less resolved for large $h$ where the trajectories start at a much later time, but that all discretizations converge towards the same long time behavior.

\subsection{Comparison of mesoscopic and macroscopic simulations}
\begin{figure}[H]
\subfigure[]{\includegraphics[width=.31\textwidth]{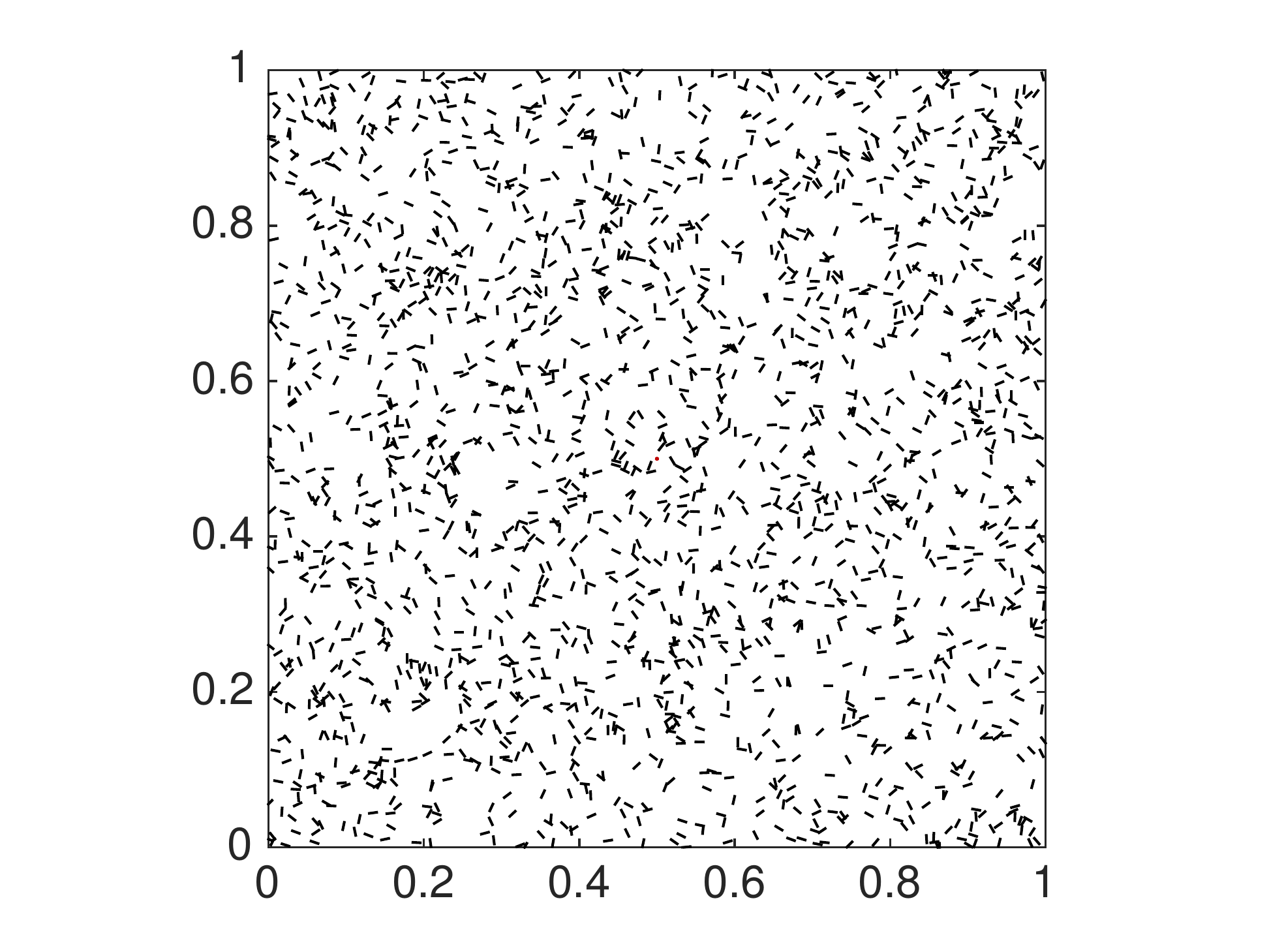} }
\subfigure[]{\includegraphics[width=.31\textwidth]{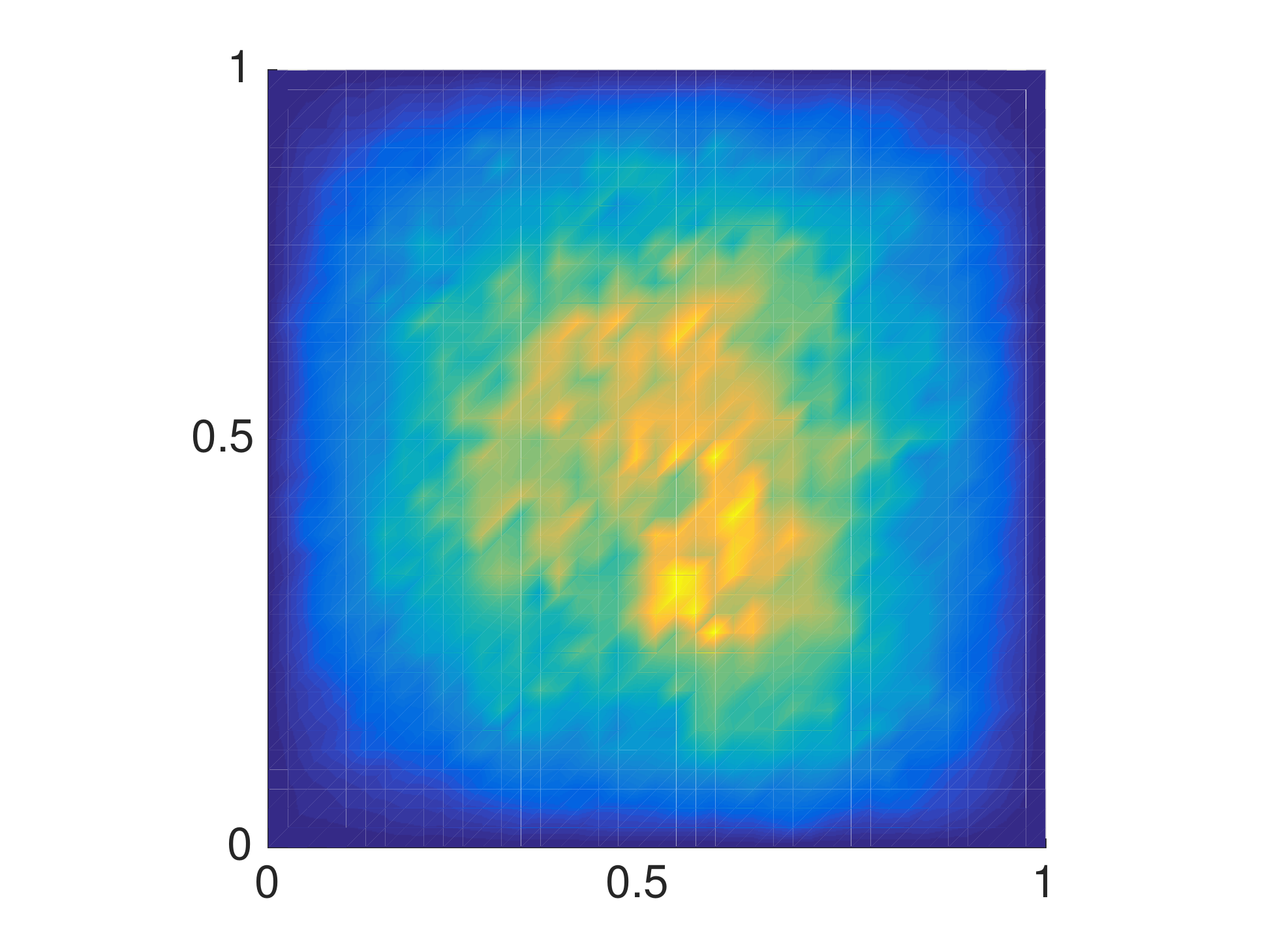} }
\subfigure[]{\includegraphics[width=.31\textwidth]{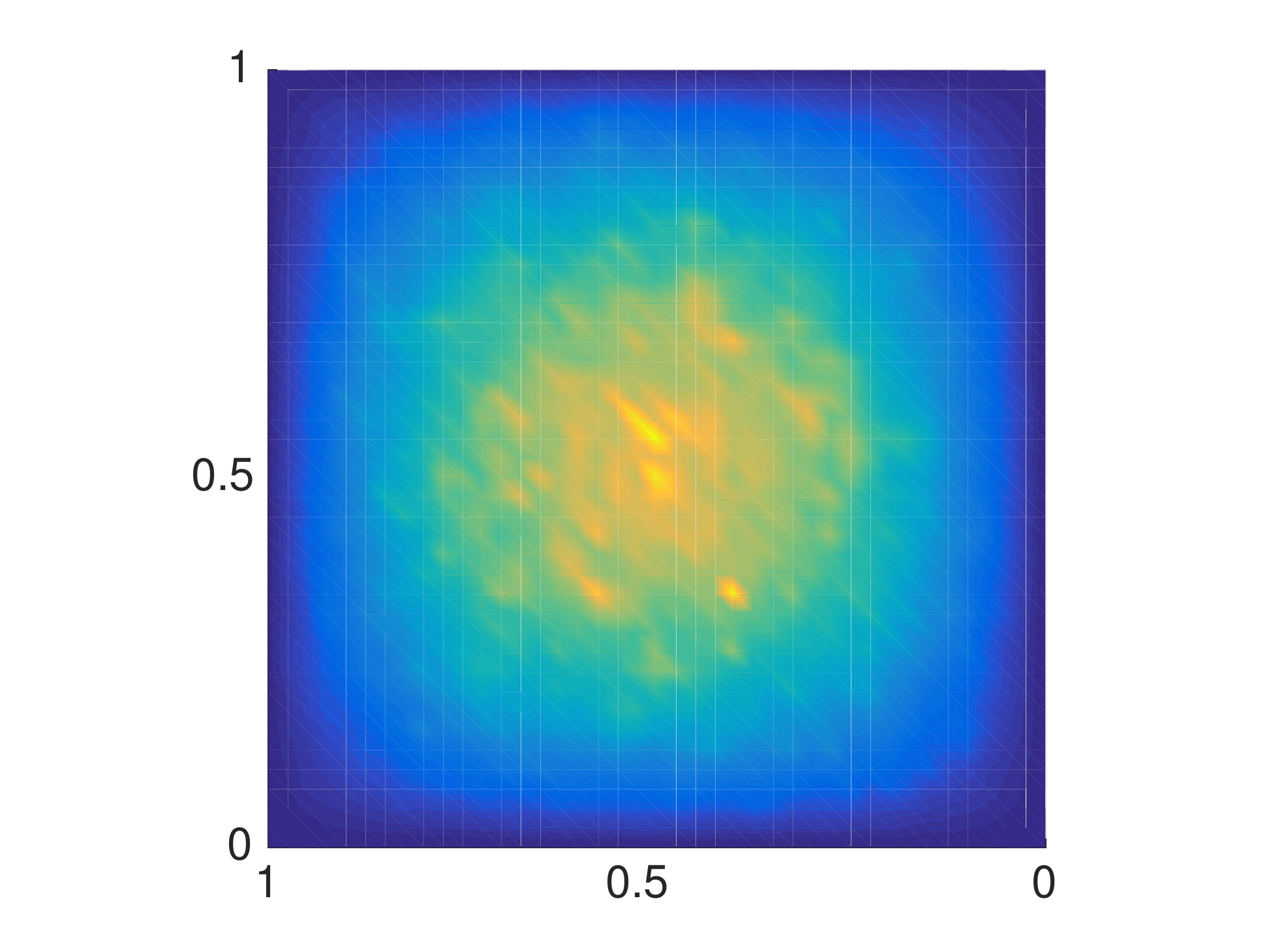} }\\
\subfigure[]{\includegraphics[width=.31\textwidth]{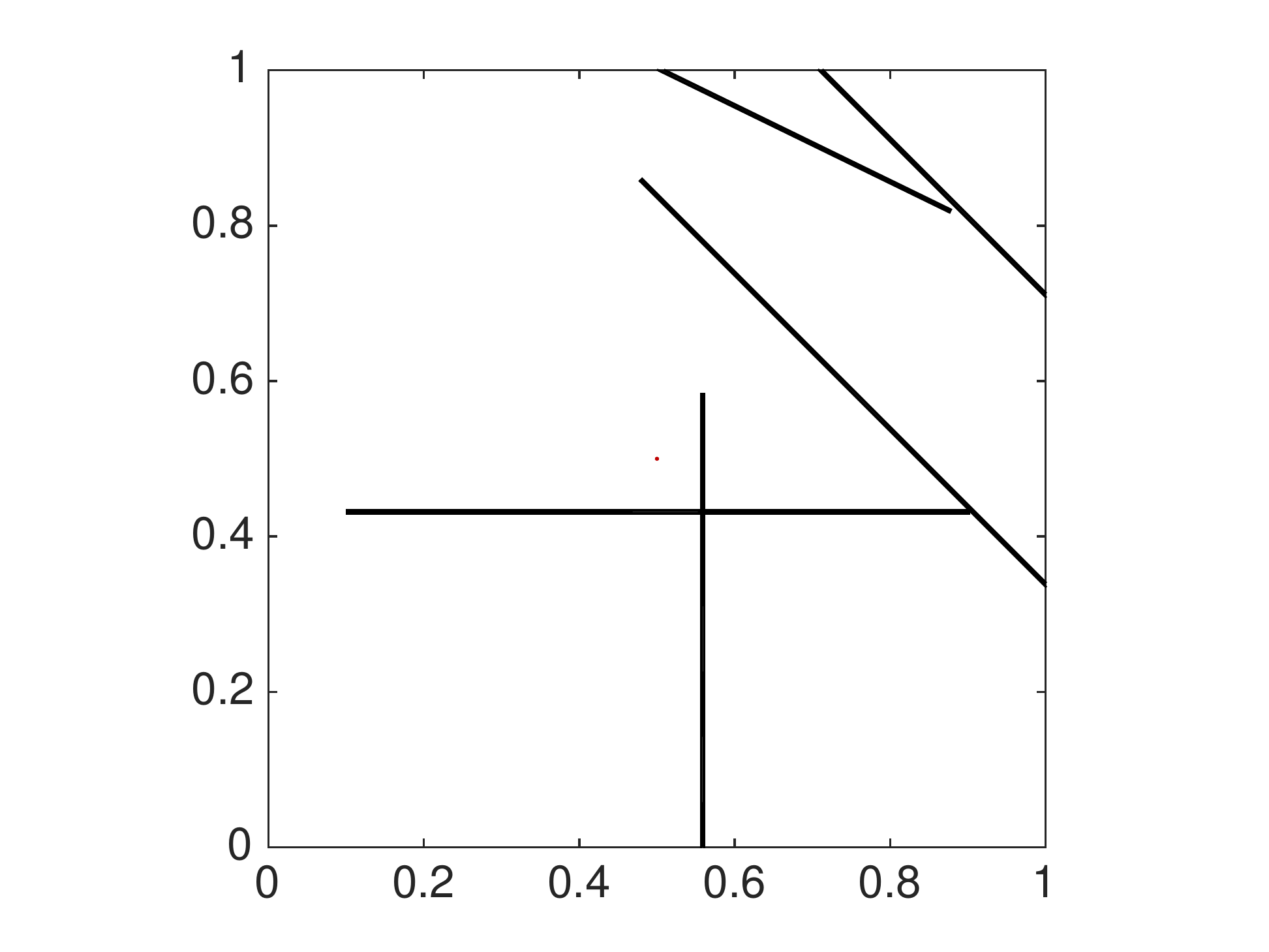} }
\subfigure[]{\includegraphics[width=.31\textwidth]{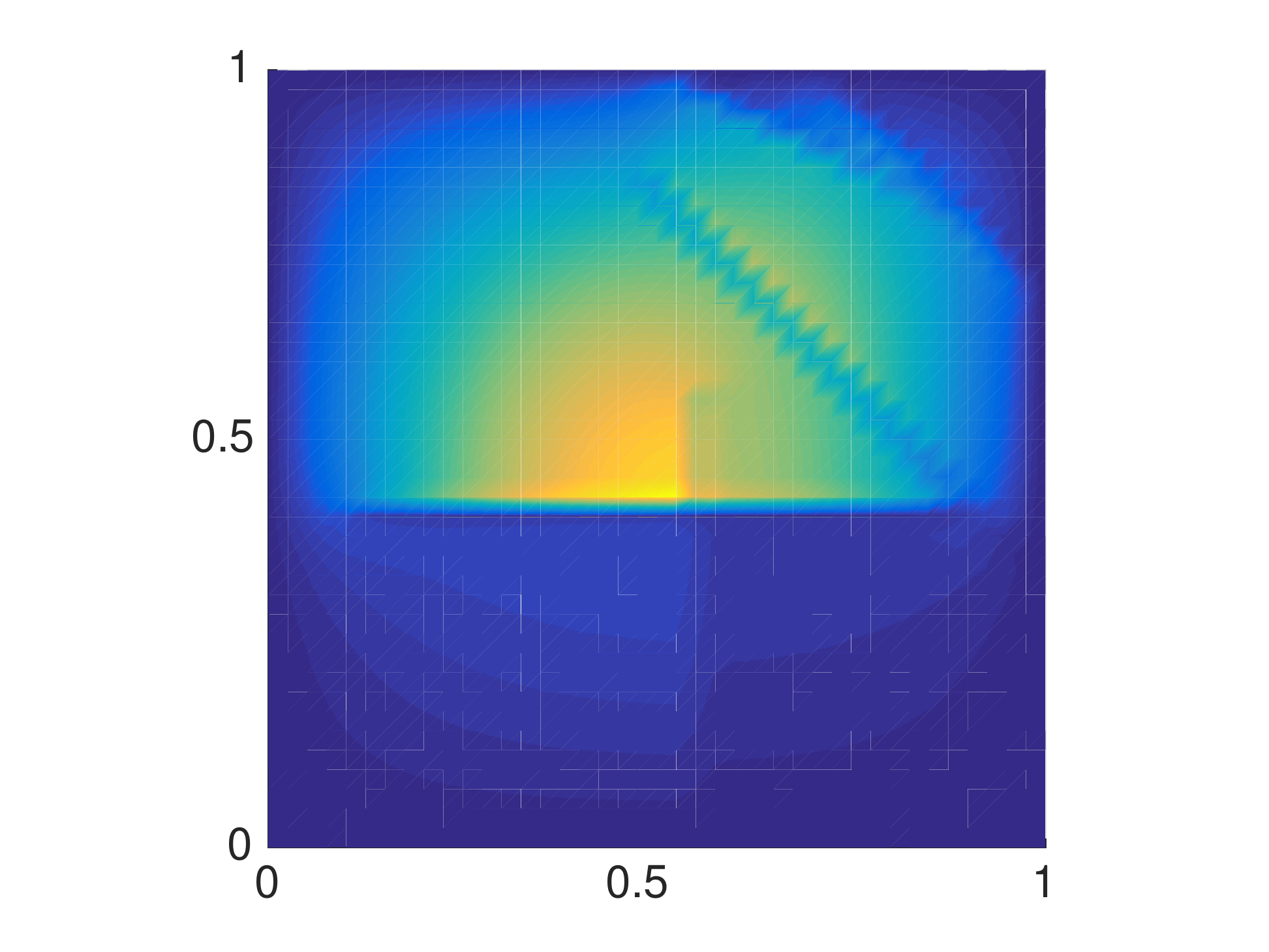} }
\subfigure[]{\includegraphics[width=.31\textwidth]{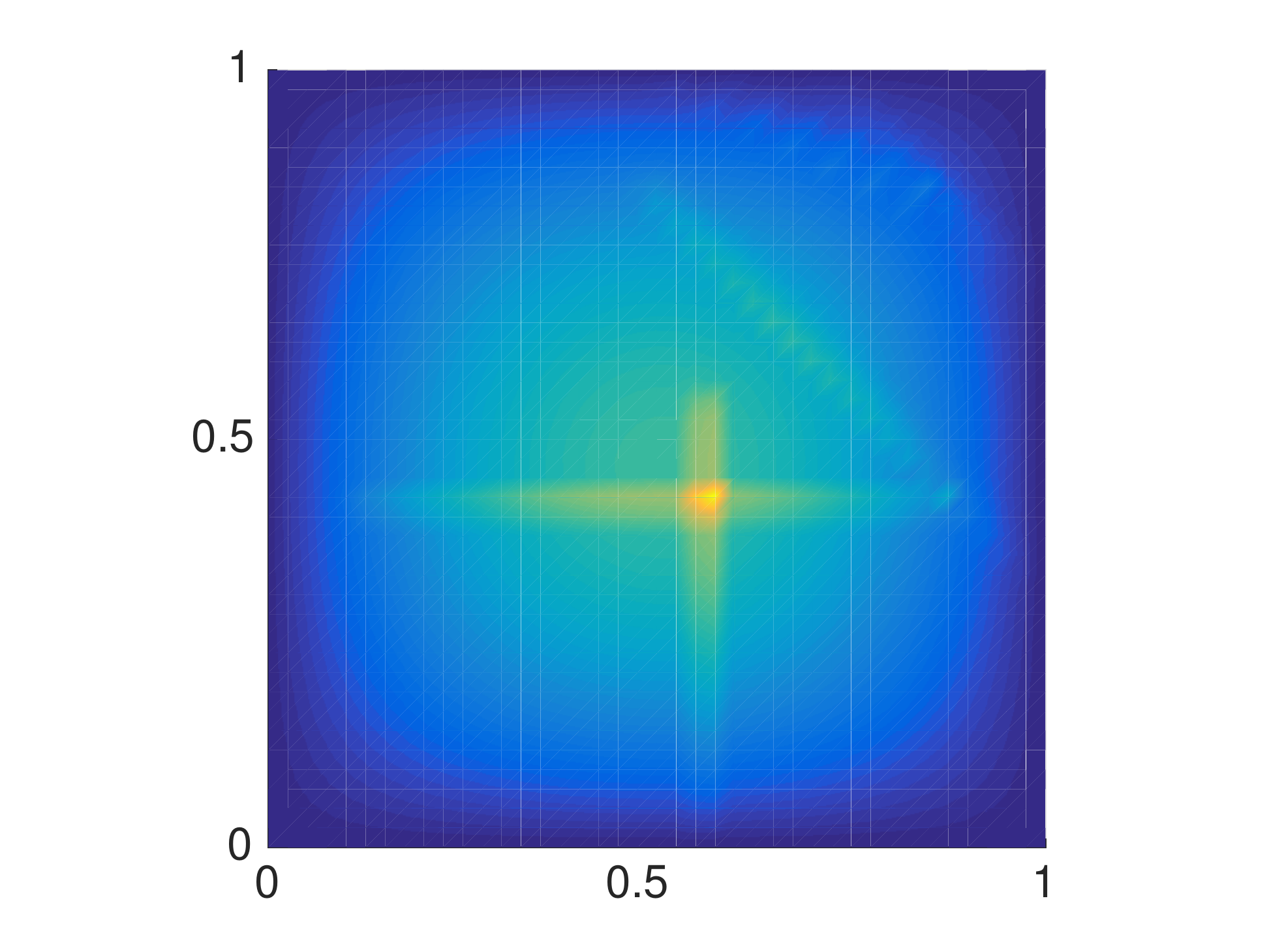} }
\caption{Solution to \eqref{eq:pODE} with different sized rectangular crowders and a moving molecule with radius $r = 1e-3$. Results are shown for $t=0.5$ with homogenous Dirichlet boundary conditions. (a) Rectangles of size $5e-4\times 1e-2$ and $\phi=0.01$. (d) 5 rectangles of size $0.004\times 0.8$. (b) $\&$ (e) Mesoscopic simulation. (c) $\&$ (f) Symmetrized macroscopic simulation.}
\label{fig:MesoMacro}
\end{figure}

The MSD is only one quantity of interest to examine, but since it is a mean not all features are captured and we will now compare the distributions of molecules resulting from either a mesoscopic or macroscopic simulation.
Again, we discretize the square $[0,1]\times[0,1]$ with homogeneous Dirichlet boundary conditions into 41 nodes in each direction and let molecules start diffusing in $(0.5,0.5)$ in an environment with rectangles of different sizes.
We solve \eqref{eq:pODE} once with the mesoscopic $\fatD$ and once with $\tilde{\fatD}$, where the off-diagonal elements are all equal to $\lambda_{ii}/4$, corresponding to a finite difference approximation of the macroscopic equation with the space dependent diffusion constant $\gamma(\fatx)$ derived from the $\lambda_{ii}$s.
In Fig.~\ref{fig:MesoMacro} the macroscopic model agrees with the mesoscopic results for small and evenly distributed crowding molecules , whereas for long barriers only the mesoscopic approach simulates the expected behavior.
This is due to the symmetrization of $\tilde{\fatD}$, so that only $\fatD$ can capture the asymmetric diffusion close to the barriers.
The diagonal barriers are not completely impermeable in the mesoscopic model since a small part of the boundary $\partial\omega_{ij}$ remains.

\subsection{Reaction rates}\label{sec:Reactions}
\begin{figure}[H]
\centering
\subfigure[]{\includegraphics[width=.25\textwidth]{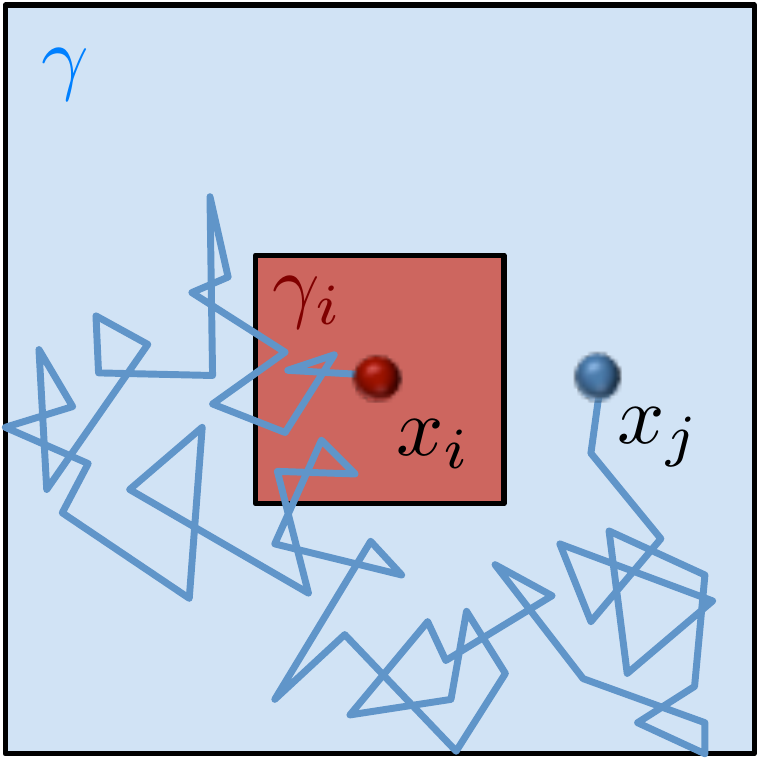} }
\subfigure[]{\includegraphics[width=.34\textwidth]{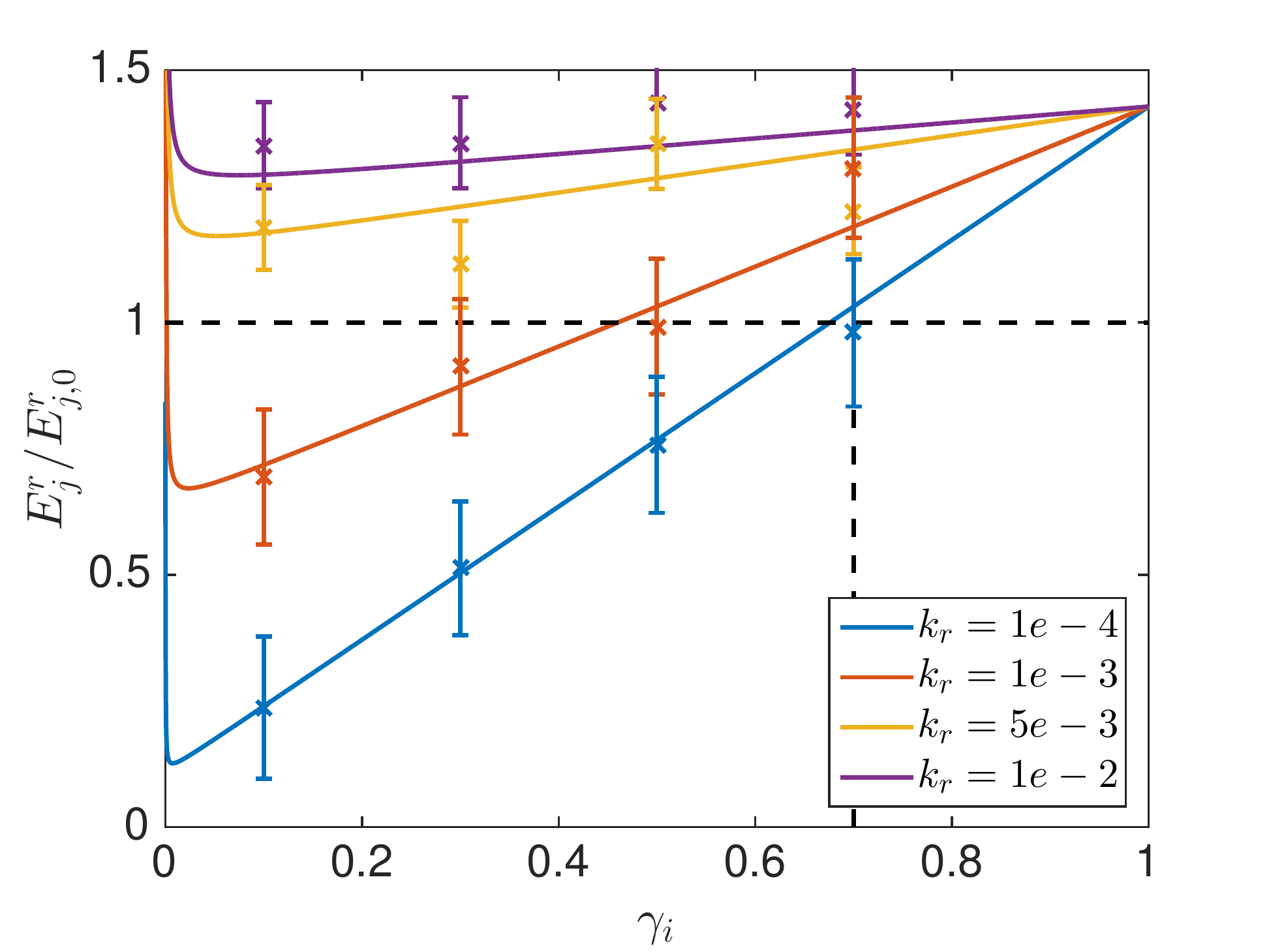} }
\subfigure[]{\includegraphics[width=.34\textwidth]{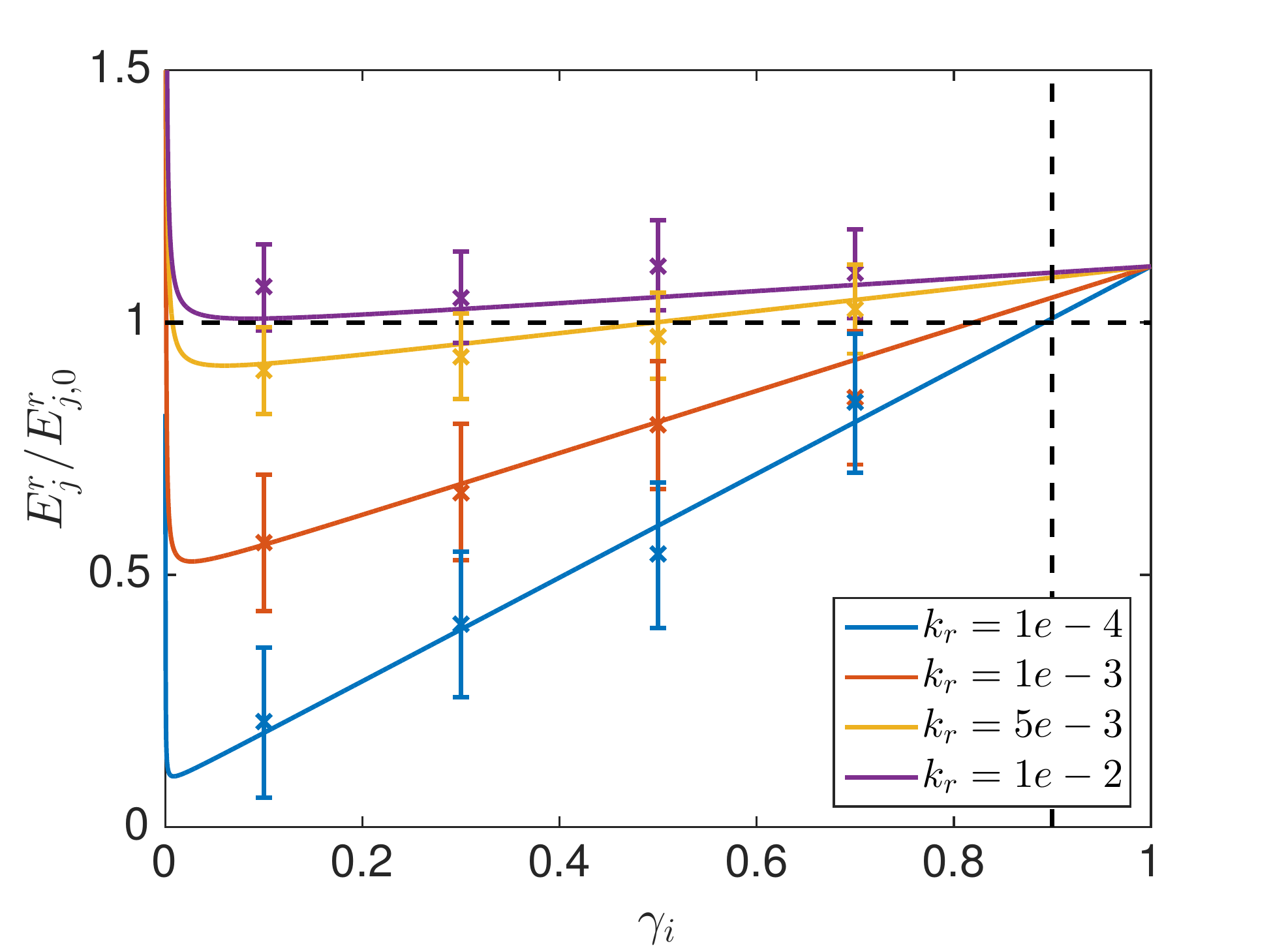} }
\caption{(a) Experimental setting where a $B$ molecule starts diffusing in $\fatx_j=(0.7,0.5,0.5)$ and reacts with $A$ that is confined to voxel $\calV_i$ with $\fatx_i=(0.5,0.5,0.5)$. Due to an uneven distribution of crowders we assume that the diffusion rate is $\gamma_i$ inside $\calV_i$ and $\gamma$ (vertical dashed line) in the rest of the domain. The time it takes to react in this crowded environment $\fatE^r_j$ is compared to that in an uncrowded environment $\fatE^r_{j,0}$ with $\gamma_0=1$. (b) $\gamma = 0.7$. (c) $\gamma=0.9$.}
\label{fig:Ereact}
\end{figure}

Due to the reduction of diffusion in a crowded environment and \eqref{eq:ReactionRate} the overall reaction rate is decreased.
It has, however, been shown \cite{Grasberger1986,Schnell2004}, that protein associations can also be enhanced.
We examine the mean time $\mathbb{E}^{r}_j$ until the bimolecular reaction \eqref{eq:Reaction}
\begin{equation}
A+B\xrightarrow{k_r} C
\label{eq:Reaction}
\end{equation}
happens, where reactant $A$ is confined to voxel $\calV_i$, and molecule $B$ starts diffusing in voxel $\calV_j$ at time $t=0$, see Fig.~\ref{fig:Ereact}(a).
Due to molecular crowding we assume a simplified space dependent diffusion map with $\gamma_i<\gamma_0$ inside $\calV_i$ and $\gamma<\gamma_0$ in the rest of the domain.
With $k_i$ given by \eqref{eq:ReactionRate} and $\lambda_{ii}$ by \eqref{eq:Lambdaii} we can use conditioning on the first step to compute the expected time until the reaction happens:
\begin{eqnarray}
\mathbb{E}^r_i &=& \frac{k_i}{k_i+\lambda_{ii}}\frac{1}{k_i+\lambda_{ii}}+\frac{\lambda_{ii}}{k_i+\lambda_{ii}}\left[\frac{1}{k_i+\lambda_{ii}}+\sum_m\theta_{im}E(\fatx_m)+\mathbb{E}^r_i\right]\label{eq:Eri}\\
\mathbb{E}^r_j &=& E(\fatx_j)+\frac{h^2+4\gamma_i\sum_m\theta_{im}E(\fatx_m)}{h^2k_i},
\end{eqnarray}
where $E(\fatx_m)$ is the expected time it takes a molecule located in the neighboring voxel $\calV_m$ to jump into voxel $\calV_i$ and can be computed by solving 
\begin{eqnarray}
\fatp_t(t) &=& (\fatD-\fatK_i)\fatp(t),\quad\fatp(0)=\fatp_0,\label{eq:Ereact1}\\
E(\fatx_m) &=& \int_0^\infty\sum_{k=1}^N|\calV_k|p_{k}(t)dt,\label{eq:Ereact2}
\end{eqnarray}
where $\fatK_i$ models a sink at node $\fatx_i$ and is the zero matrix except for $K_{i,i}=10^9$ and $\fatp_0$ is the zero vector except for $p_{0,m}=1/h^3$, see \cite{Meinecke2016-2} for a derivation.
We solve these equations numerically in 3D for a cube with length $L=1$ and a uniform discretization with $h=0.1$ in space and reflecting boundary conditions. The voxel $\calV_i$, where the reaction happens is chosen to be the center voxel, such that $E(\fatx_m)$ are equal for all 6 neighbors. The diffusing molecule $B$ starts in $\fatx_j=(0.7,0.5,0.5)$.
In Fig.~\ref{fig:Ereact} we compare the mean binding time in the crowded $\mathbb{E}_j^r$ environment with different $\gamma_i$ and $\gamma$ to the time $\mathbb{E}^r_{j,0}$ it takes to react in a dilute solution where $\gamma(\fatx)=\gamma_0=1$.
The data points with scaled error bars ($\pm(\frac{\sigma}{\sqrt{M}\mathbb{E}^r_j}+\frac{\sigma_0}{\sqrt{M}\mathbb{E}^r_{j,0}})$) are from a SSA simulation of the reaction-diffusion process with $M=200$ trajectories for $k_r=1e-4$ and $k_r=1e-3$ and $M=500$ for $k_r=5e-3$ and $k_r=1e-2$.
An overall slower diffusion rate $\gamma<\gamma_0$ as a result of obstacles reduces the rate of bimoleculear reactions \eqref{eq:ReactionRate} in each voxel.
But, due to an uneven distribution of crowding agents compartmentalization with locally differing diffusion rates can occur inside the cells.
In Fig.~\ref{fig:Ereact} the overall reaction rate can be increased for a compartmentalization where $\gamma_i$ is smaller than $\gamma$, despite the locally slower reaction.
Because once the diffusing molecule enters $\calV_i$ it escapes slower and hence gets trapped close to its reaction partner, which increases the chance of collisions.
Like this, cells can boost their efficiency by locating important reaction complexes in areas of slower diffusion, which will be especially productive for reaction cascades where intermediate products are already produced inside the compartment and have a low chance of escaping before being processed further.
In the limit when $\gamma_i\to0$ the binding time goes towards infinity, since then both reactants are immobile inside $\calV_i$ and never collide.


\section{Conclusion}
We have presented a multiscale framework to model diffusion and reactions in a crowded environment, which is an important feature for realistic simulations inside living cells and on their membranes.
First, we solve a set of PDEs on local domains resolving the microscopic positions and shapes of the crowding molecules. 
This pre-computing step is embarrassingly parallelizable and yields local first exit times, which can be transformed into the jump rates on an overlying Cartesian grid at the mesoscopic level.
We then use these local first exit times to compute a space dependent diffusion coefficient for the macroscopic diffusion equation, which corresponds to space dependent reaction rates according to the formula by Colins and Kimball.

Our approach is general in the sense that the crowding molecules can have arbitrary shapes and can be located anywhere inside the domain.
We indicate how to adapt our method to moving crowders by computing statistics which we will further explore in the future.
As the jump process is simulated on a coarse Cartesian mesh, no longer resolving the numerous crowders, the stochastic simulation is computationally much more efficient than a microscopic simulation capturing all the collisions.

In numerical experiments we foremost observe that shape and size considerably affect how strongly diffusion is impeded: small crowders have more reflective surface and hence hinder diffusion more severely than bigger obstacles, so do elongated crowders, which create long barriers.
The effect is also stronger for larger diffusing molecules than for smaller ones, since the former need bigger gaps to pass through.
This gives some new insight into how non-idealized (non-spherical) macromolecules affect the diffusion, since most existing models either assume that all particles are spheres or only consider the percentage of occupied volume.

Comparing the mesoscopic and macroscopic models for diffusion in the crowded environment we note that the former captures the asymmetries created by long barriers better and that they both behave similarly for small crowding molecules compared to the grid size.

The space dependent diffusion rate can be interpreted as a compartmentalization effect, which has been observed in cells.
In a simplified example we see that reactions located inside a compartment with high crowding/low diffusivity can be enhanced since the reaction partners reside longer in the vicinity of each other.
Hence, the concentration of reaction complexes in an area with slow diffusion as compared to the rest of the cytoplasm or cell membrane can increase the reaction turn-over, an effect that has been capitalized by cells through co-localization of reaction complexes and scaffolding.
Otherwise, reactions between initially distant molecules are impeded by excluded volume since it takes longer time for the reactants to find each other.

Hard sphere reflections on obstacles are not the sole cause of anomalous diffusion \cite{Ridgway2008}, but there are other interactions between macromolecules, such as transient binding or electrostatic repulsion, which have been modeled by a continous time random walk \cite{Barkai2012,Schulz2014} and fractional or multifractional Brownian motion \cite{Marquez-Lago2012}.
We can include these types of interactions, by modifying boundary conditions on the crowders from reflecting to partially absorbing or adding potential barriers.

\section*{Acknoledgements}
This work was supported by the Swedish Research Council grant 621-2001-3148 and the NIH grant for StochSS with number 1R01EB014877-01. The author would like to thank the Computational Systems Biology group at Uppsala University for fruitful discussions and Markus Eriksson for the Smoldyn simulations.

\bibliographystyle{plain}
\bibliography{mesomicro,Crowding2}

\newcommand{\noopsort}[1]{}
\begin{thebibliography}{10}

\bibitem{AnAdBrAr10}
S.~S. Andrews, N.~J. Addy, R.~Brent, and A.~P. Arkin.
\newblock Detailed simulations of cell biology with {S}moldyn 2.1.
\newblock {\em PLoS Comput. Biol.}, 6(3):e1000705, 2010.

\bibitem{Andrews2004}
S~S Andrews and D~Bray.
\newblock {Stochastic simulation of chemical reactions with spatial resolution
  and single molecule detail.}
\newblock {\em Phys. Biol.}, 1(3-4):137--151, 2004.

\bibitem{Barkai2012}
Eli Barkai, Yuval Garini, and Ralf Metzler.
\newblock {Strange kinetics of single molecules in living cells}.
\newblock {\em Phys. Today}, 65(8):29--35, 2012.

\bibitem{Ben-Avraham2000}
Daniel Ben-Avraham and Shlomo Havlin.
\newblock {\em Diffusion and reactions in fractals and disordered systems}.
\newblock Cambridge University Press, 2000.

\bibitem{Berry2002}
Hugues Berry.
\newblock {Monte carlo simulations of enzyme reactions in two dimensions:
  fractal kinetics and spatial segregation.}
\newblock {\em Biophys. J.}, 83(4):1891--1901, 2002.

\bibitem{Blanc2015}
Emilie Blanc, Stefan Engblom, Andreas Hellander, and Per L{\"{o}}tstedt.
\newblock {Mesoscopic modeling of stochastic reaction-diffusion kinetics in the
  subdiffusive regime}.
\newblock pages 1--30, 2015.

\bibitem{Brown2014}
D.~L. {Brown} and D.~{Peterseim}.
\newblock {A Multiscale Method for Porous Microstructures}.
\newblock {\em ArXiv e-prints}, November 2014.

\bibitem{CaoGilPet1}
Y.~Cao, D.~T. Gillespie, and L.~R. Petzold.
\newblock The slow-scale stochastic simulation algorithm.
\newblock {\em J.~Chem.~Phys.}, 122:014116, 2005.

\bibitem{Cianci2015}
Claudia Cianci, Stephen Smith, and Ramon Grima.
\newblock {Molecular finite-size effects in stochastic models of equilibrium
  chemical systems}.
\newblock {\em J. Chem. Phys.}, 084101(144):1--35, 2016.

\bibitem{CoKi}
F.~C. Collins and G.~E. Kimball.
\newblock Diffusion-controlled reaction rates.
\newblock {\em J. Colloid. Sci.}, 4:425--437, 1949.

\bibitem{DiRienzo2014}
Carmine {Di Rienzo}, Vincenzo Piazza, Enrico Gratton, Fabio Beltram, and
  Francesco Cardarelli.
\newblock {Probing short-range protein Brownian motion in the cytoplasm of
  living cells}.
\newblock {\em Nat. Commun.}, 5:5891, 2014.

\bibitem{DBOGSK}
A.~Donev, V.~V. Bulatov, T.~Oppelstrup, G.~H. Gilmer, B.~Sadigh, and M.~H.
  Kalos.
\newblock A first-passage kinetic {M}onte {C}arlo algorithm for complex
  diffusion-reaction systems.
\newblock {\em J.~Comput.~Phys.}, 229:3214--3236, 2010.

\bibitem{Donev2010a}
Aleksandar Donev, Vasily~V. Bulatov, Tomas Oppelstrup, George~H. Gilmer, Babak
  Sadigh, and Malvin~H. Kalos.
\newblock {A First-Passage Kinetic Monte Carlo algorithm for complex
  diffusion–reaction systems}.
\newblock {\em J. Comput. Phys.}, 229(9):3214--3236, may 2010.

\bibitem{URDME}
B.~Drawert, S.~Engblom, and A.~Hellander.
\newblock {URDME}: a modular framework for stochastic simulation of
  reaction-transport processes in complex geometries.
\newblock {\em BMC Syst. Biol.}, 6:76, 2012.

\bibitem{ElEh04}
J.~Elf and M.~Ehrenberg.
\newblock Spontaneous separation of bi-stable biochemical systems into spatial
  domains of opposite phases.
\newblock {\em Syst. Biol.}, 1:230--236, 2004.

\bibitem{Ellery2015}
Adam~J Ellery, Ruth~E Baker, and Matthew~J Simpson.
\newblock {Calculating the Fickian diffusivity for a lattice-based random walk
  with agents and obstacles of different shapes and sizes}.
\newblock {\em Phys. Biol.}, 12(6):066010, 2015.

\bibitem{Ellis2001}
R.~J. Ellis.
\newblock {Macromolecular crowding: An important but neglected aspect of the
  intracellular environment}.
\newblock {\em Curr. Opin. Struct. Biol.}, 11(1):114--119, 2001.

\bibitem{ELSS}
M.~B. Elowitz, A.~J. Levine, E.~D. Siggia, and P.~S. Swain.
\newblock Stochastic gene expression in a single cell.
\newblock {\em Science}, 297:1183--1186, 2002.

\bibitem{EnFeHeLo}
S.~Engblom, L.~Ferm, A.~Hellander, and P.~L\"{o}tstedt.
\newblock Simulation of stochastic reaction-diffusion processes on unstructured
  meshes.
\newblock {\em SIAM J.~Sci.~Comput.}, 31:1774--1797, 2009.

\bibitem{Fanelli2013}
D~Fanelli, a~J McKane, G~Pompili, B~Tiribilli, M~Vassalli, and T~Biancalani.
\newblock {Diffusion of two molecular species in a crowded environment: theory
  and experiments.}
\newblock {\em Phys. Biol.}, 10(4):045008, 2013.

\bibitem{Fanelli2010}
Duccio Fanelli and Alan~J. McKane.
\newblock {Diffusion in a crowded environment}.
\newblock {\em Phys. Rev. E - Stat. Nonlinear, Soft Matter Phys.}, 82(2):1--4,
  2010.

\bibitem{Fange2010}
David Fange, Otto~G Berg, Paul Sj{\"{o}}berg, and Johan Elf.
\newblock {Stochastic reaction-diffusion kinetics in the microscopic limit.}
\newblock {\em Proc. Natl. Acad. Sci. U. S. A.}, 107(46):19820--5, nov 2010.

\bibitem{Gardiner}
C.~W. Gardiner.
\newblock {\em Handbook of Stochastic Methods}.
\newblock Springer Series in Synergetics. Springer-Verlag, Berlin, 3rd edition,
  2004.

\bibitem{GaMcWaMa}
C.~W. Gardiner, K.~J. McNeil, D.~F. Walls, and I.~S. Matheson.
\newblock Correlations in stochastic theories of chemical reactions.
\newblock {\em J. Stat. Phys.}, 14(4):307--331, 1976.

\bibitem{GibsonBruck}
M.~A. Gibson and J.~Bruck.
\newblock Efficient exact stochastic simulation of chemical systems with many
  species and many channels.
\newblock {\em J.~Phys.~Chem.}, 104(9):1876--1889, 2000.

\bibitem{gillespie}
D.~T. Gillespie.
\newblock A general method for numerically simulating the stochastic time
  evolution of coupled chemical reactions.
\newblock {\em J.~Comput.~Phys.}, 22(4):403--434, 1976.

\bibitem{Gillespie2013}
Daniel~T. Gillespie, Andreas Hellander, and Linda~R. Petzold.
\newblock {Perspective: Stochastic algorithms for chemical kinetics}.
\newblock {\em J. Chem. Phys.}, 138(17), 2013.

\bibitem{Grasberger1986}
B~Grasberger, a~P Minton, C~DeLisi, and H~Metzger.
\newblock {Interaction between proteins localized in membranes.}
\newblock {\em Proc. Natl. Acad. Sci. U. S. A.}, 83(17):6258--6262, 1986.

\bibitem{Grima2010}
R.~Grima.
\newblock {Intrinsic biochemical noise in crowded intracellular conditions}.
\newblock {\em J. Chem. Phys.}, 132(18), 2010.

\bibitem{Grima2006}
R.~Grima and S.~Schnell.
\newblock {A systematic investigation of the rate laws valid in intracellular
  environments}.
\newblock {\em Biophys. Chem.}, 124(1):1--10, 2006.

\bibitem{Grima2007}
Ramon Grima and Santiago Schnell.
\newblock {A mesoscopic simulation approach for modeling intracellular
  reactions}.
\newblock {\em J. Stat. Phys.}, 128(1-2):139--164, 2007.

\bibitem{Hall2003}
Damien Hall and Allen~P. Minton.
\newblock {Macromolecular crowding: Qualitative and semiquantitative successes,
  quantitative challenges}.
\newblock {\em Biochim. Biophys. Acta - Proteins Proteomics}, 1649(2):127--139,
  2003.

\bibitem{Hansen2015}
Maike M.~K. Hansen, Lenny H.~H. Meijer, Evan Spruijt, Roel J.~M. Maas,
  Marta~Ventosa Rosquelles, Joost Groen, Hans~A. Heus, and Wilhelm T.~S. Huck.
\newblock {Macromolecular crowding creates heterogeneous environments of gene
  expression in picolitre droplets}.
\newblock {\em Nat. Nanotechnol.}, 11(October):1--8, 2015.

\bibitem{HFE}
J.~Hattne, D.~Fange, and J.~Elf.
\newblock Stochastic reaction-diffusion simulation with {M}eso{RD}.
\newblock {\em Bioinformatics}, 21:2923--2924, 2005.

\bibitem{Havlin2002}
Shlomo Havlin and Daniel Ben-Avraham.
\newblock {Diffusion in disordered media}.
\newblock {\em Adv. Phys.}, 51(1):187--292, 2002.

\bibitem{Hellander2012}
Stefan Hellander, Andreas Hellander, and Linda Petzold.
\newblock {Reaction-diffusion master equation in the microscopic limit}.
\newblock {\em Phys. Rev. E - Stat. Nonlinear, Soft Matter Phys.}, 85(4):1--5,
  2012.

\bibitem{Hellander2015}
Stefan Hellander, Andreas Hellander, and Linda Petzold.
\newblock {Reaction rates for mesoscopic reaction-diffusion kinetics}.
\newblock {\em Phys. Rev. E}, 91(2), 2015.

\bibitem{STEPS}
I.~Hepburn, W.~Chen, S.~Wils, and E.~De Schutter.
\newblock {STEPS}: efficient simulation of stochastic reaction-diffusion models
  in realistic morphologies.
\newblock {\em BMC Syst. Biol.}, 6:36, 2012.

\bibitem{Hrabe2004}
Jan Hrabe, Sabina Hrabetov{\'{a}}, and Karel Segeth.
\newblock {A model of effective diffusion and tortuosity in the extracellular
  space of the brain.}
\newblock {\em Biophys. J.}, 87(3):1606--1617, 2004.

\bibitem{IsaacsonPeskin}
S.~A. Isaacson and C.~S. Peskin.
\newblock Incorporating diffusion in complex geometries into stochastic
  chemical kinetics simulations.
\newblock {\em SIAM J. Sci. Comput.}, 28(1):47--74, 2006.

\bibitem{Isaacson2009}
Samuel~A Isaacson.
\newblock The reaction-diffusion master equation as an asymptotic approximation
  of diffusion to a small target.
\newblock {\em SIAM Journal on Applied Mathematics}, 70(1):77--111, 2009.

\bibitem{Jin2007}
Songwan Jin and a.~S. Verkman.
\newblock {Single particle tracking of complex diffusion in membranes:
  Simulation and detection of barrier, raft, and interaction phenomena}.
\newblock {\em J. Phys. Chem. B}, 111(14):3625--3632, 2007.

\bibitem{MCell08}
R.~A. Kerr, T.~M. Bartol, B.~Kaminsky, M.~Dittrich, J.-C.~J. Chang, S.~B.
  Baden, T.~J. Sejnowski, and J.~R. Stiles.
\newblock Fast {M}onte {C}arlo simulation methods for biological
  reaction-diffusion systems in solution and on surfaces.
\newblock {\em SIAM J. Sci. Comput.}, 30(6):3126--3149, 2008.

\bibitem{Krapf2015}
Diego Krapf.
\newblock {\em {Mechanisms Underlying Anomalous Diffusion in the Plasma
  Membrane}}, volume~75.
\newblock Elsevier Ltd, 2015.

\bibitem{Landman2011}
Kerry~a. Landman and Anthony~E. Fernando.
\newblock {Myopic random walkers and exclusion processes: Single and
  multispecies}.
\newblock {\em Phys. A Stat. Mech. its Appl.}, 390(21-22):3742--3753, 2011.

\bibitem{Lee2008}
Byoungkoo Lee, Philip~R. LeDuc, and Russell Schwartz.
\newblock {Stochastic off-lattice modeling of molecular self-assembly in
  crowded environments by Green’s function reaction dynamics}.
\newblock {\em Phys. Rev. E}, 78(3):031911, 2008.

\bibitem{Lotstedt2015}
Per L{\"{o}}tstedt and Lina Meinecke.
\newblock {Simulation of stochastic diffusion via first exit times}.
\newblock {\em J. Comput. Phys.}, 300:862--886, nov 2015.

\bibitem{Luby-Phelps2000}
K~Luby-Phelps.
\newblock {Cytoarchitecture and physical properties of cytoplasm: volume,
  viscosity, diffusion, intracellular surface area.}, 2000.

\bibitem{Malqvist2014}
Axel M{\aa}lqvist and Daniel Peterseim.
\newblock {Localization of elliptic multiscale problems}.
\newblock {\em Math. Comput.}, 83(290):2583--2603, 2014.

\bibitem{Marquez-Lago2012}
T.T. Marquez-Lago, a.~Leier, and K.~Burrage.
\newblock {Anomalous diffusion and multifractional Brownian motion: simulating
  molecular crowding and physical obstacles in systems biology}.
\newblock {\em IET Syst. Biol.}, 6(4):134, 2012.

\bibitem{MAA1}
H.~H. McAdams and A.~Arkin.
\newblock Stochastic mechanisms in gene expression.
\newblock {\em Proc. Natl. Acad. Sci. USA}, 94:814--819, 1997.

\bibitem{McQuarrie}
D.~A. McQuarrie.
\newblock Stochastic approach to chemical kinetics.
\newblock {\em J. Appl. Prob.}, 4:413--478, 1967.

\bibitem{Medalia2002}
Ohad Medalia, Igor Weber, Achilleas~S Frangakis, Daniela Nicastro, Gunther
  Gerisch, and Wolfgang Baumeister.
\newblock {Macromolecular architecture in eukaryotic cells visualized by
  cryoelectron tomography.}
\newblock {\em Science}, 298(2002):1209--1213, 2002.

\bibitem{Meinecke2016-2}
Lina Meinecke, Stefan Engblom, Andreas Hellander, and Per L\"{o}tstedt.
\newblock {Analysis and design of jump coefficients in discrete stochastic
  diffusion models.}
\newblock {\em SIAM J. Sci. Comput.}, 38(1):A55--A83, 2016.

\bibitem{Meinecke2016}
Lina Meinecke and Per L{\"{o}}tstedt.
\newblock {Stochastic diffusion processes on Cartesian meshes}.
\newblock {\em J. Comput. Appl. Math.}, 294:1--11, mar 2016.

\bibitem{Met}
R.~Metzler.
\newblock The future is noisy: {T}he role of spatial fluctuations in genetic
  switching.
\newblock {\em Phys. Rev. Lett.}, 87:068103, 2001.

\bibitem{Mommer2009}
Mario~S Mommer and Dirk Lebiedz.
\newblock Modeling subdiffusion using reaction diffusion systems.
\newblock {\em SIAM Journal on Applied Mathematics}, 70(1):112--132, 2009.

\bibitem{MunskyNeuertOuden}
B.~Munsky, G.~Neuert, and A.~van Oudenaarden.
\newblock Using gene expression noise to understand gene regulation.
\newblock {\em Science}, 336(6078):183--187, 2012.

\bibitem{Muramatsu1988}
N~Muramatsu and a~P Minton.
\newblock {Tracer diffusion of globular proteins in concentrated protein
  solutions.}
\newblock {\em Proc. Natl. Acad. Sci. U. S. A.}, 85(9):2984--2988, 1988.

\bibitem{Oksendal}
B.~{\O}ksendal.
\newblock {\em Stochastic Differential Equations}.
\newblock Springer, Berlin, 6th edition, 2003.

\bibitem{OBDKGS}
T.~Oppelstrup, V.~V. Bulatov, A.~Donev, M.~H. Kalos, G.~H. Gilmer, and
  B.~Sadigh.
\newblock First-passage kinetic {M}onte {C}arlo method.
\newblock {\em Phys.~Rev.~E}, 80:066701, 2009.

\bibitem{Penington2011}
Catherine~J. Penington, Barry~D. Hughes, and Kerry~a. Landman.
\newblock {Building macroscale models from microscale probabilistic models: A
  general probabilistic approach for nonlinear diffusion and multispecies
  phenomena}.
\newblock {\em Phys. Rev. E}, 84(4):041120, 2011.

\bibitem{Phillips2009}
Rob Phillips, Jane Kondev, and Julie Theriot.
\newblock {\em Physical Biology of the Cell}.
\newblock Garland Science, Taylor \& Francis Group, New York, November 2008.

\bibitem{RajOuden}
A.~Raj and A.~van Oudenaarden.
\newblock Nature, nurture, or chance: {S}tochastic gene expression and its
  consequences.
\newblock {\em Cell}, 135(2):216--226, 2008.

\bibitem{Redner}
S.~Redner.
\newblock {\em A Guide to First-Passage Processes}.
\newblock Cambridge University Press, Cambridge, 2001.

\bibitem{Ridgway2008}
Douglas Ridgway, Gordon Broderick, Ana Lopez-Campistrous, Melania Ru'aini,
  Philip Winter, Matthew Hamilton, Pierre Boulanger, Andriy Kovalenko, and
  Michael~J Ellison.
\newblock {Coarse-grained molecular simulation of diffusion and reaction
  kinetics in a crowded virtual cytoplasm.}
\newblock {\em Biophys. J.}, 94(10):3748--3759, 2008.

\bibitem{Roberts2013}
Elijah Roberts, John~E. Stone, and Zaida Luthey-Schulten.
\newblock {Lattice microbes: High-performance stochastic simulation method for
  the reaction-diffusion master equation}.
\newblock {\em J. Comput. Chem.}, 34(3):245--255, 2013.

\bibitem{Schnell2004}
S.~Schnell and T.~E. Turner.
\newblock {Reaction kinetics in intracellular environments with macromolecular
  crowding: Simulations and rate laws}.
\newblock {\em Prog. Biophys. Mol. Biol.}, 85(2-3):235--260, 2004.

\bibitem{Schoneberg2014}
Johannes Sch{\"{o}}neberg, Alexander Ullrich, and Frank No{\'{e}}.
\newblock {Simulation tools for particle-based reaction-diffusion dynamics in
  continuous space.}
\newblock {\em BMC Biophys.}, 7(1):11, 2014.

\bibitem{Schulz2014}
Johannes~H. P. Schulz, Eli Barkai, and Ralf Metzler.
\newblock {Aging Renewal Theory and Application to Random Walks}.
\newblock {\em Phys. Rev. X}, 4(1):011028, 2014.

\bibitem{Smith2014}
Gregory~R. Smith, Lu~Xie, Byoungkoo Lee, and Russell Schwartz.
\newblock {Applying Molecular Crowding Models to Simulations of Virus Capsid
  Assembly In Vitro}.
\newblock {\em Biophys. J.}, 106(1):310--320, 2014.

\bibitem{Swain01102002}
P.~S. Swain, M.~B. Elowitz, and E.~D. Siggia.
\newblock {Intrinsic and extrinsic contributions to stochasticity in gene
  expression}.
\newblock {\em Proc. Natl. Acad. Sci. USA}, 99(20):12795--12800, 2002.

\bibitem{Takahashi2010}
Koichi Takahashi, Sorin Tanase-Nicola, and Pieter~Rein ten Wolde.
\newblock {Spatio-temporal correlations can drastically change the response of
  a MAPK pathway.}
\newblock {\em Proc. Natl. Acad. Sci. U. S. A.}, 107(6):2473--2478, 2010.

\bibitem{Takahashi2005}
Kouichi Takahashi, Satya~Nanda {Vel Arjunan}, and Masaru Tomita.
\newblock {Space in systems biology of signaling pathways - Towards
  intracellular molecular crowding in silico}.
\newblock {\em FEBS Lett.}, 579(8):1783--1788, 2005.

\bibitem{Taylor2015}
P.~R. Taylor, C.~A. Yates, M.~J. Simpson, and R.~E. Baker.
\newblock {Reconciling transport models across scales: The role of volume
  exclusion}.
\newblock {\em Phys. Rev. E}, 92(4):040701, 2015.

\bibitem{Verkman2002}
Alan~S. Verkman.
\newblock {Solute and macromolecule diffusion in cellular aqueous
  compartments}.
\newblock {\em Trends Biochem. Sci.}, 27(1):27--33, 2002.

\bibitem{ZoWo5a}
{\noopsort{Zon}}{J.~S.~van~Zon and P.~R. ten Wolde}.
\newblock Green's-function reaction dynamics: {A} particle-based approach for
  simulating biochemical networks in time and space.
\newblock {\em J.~Chem.~Phys.}, 123:234910, 2005.

\end{thebibliography}

\end{document}